\documentclass[reprint,superscriptaddress,amsmath,amssymb,aps,floatfix]{revtex4-1}
\usepackage{graphicx}
\usepackage{dcolumn}
\usepackage{upgreek}
\usepackage{gensymb}
\usepackage{float}
\usepackage{bm}
\usepackage{xcolor}

\begin{document}


\title{Giant perpendicular magnetic anisotropy in Ir/Co/Pt multilayers}



\author{Yong-Chang Lau}
\affiliation{Department of Physics, The University of Tokyo, Tokyo 113-0033, Japan}
\affiliation{National Institute for Materials Science, Tsukuba 305-0047, Japan}

\author{Zhendong Chi}
\affiliation{Department of Physics, The University of Tokyo, Tokyo 113-0033, Japan}

\author{Tomohiro Taniguchi}
\affiliation{Spintronics Research Center, National Institute of Advanced Industrial Science and Technology, Tsukuba 305-8568, Japan}

\author{Masashi Kawaguchi}
\affiliation{Department of Physics, The University of Tokyo, Tokyo 113-0033, Japan}

\author{Goro Shibata}
\affiliation{Department of Physics, The University of Tokyo, Tokyo 113-0033, Japan}

\author{Naomi Kawamura}
\affiliation{Japan Synchrotron Radiation Research Institute (JASRI), Sayo 679-5198, Japan}

\author{Motohiro Suzuki}
\affiliation{Japan Synchrotron Radiation Research Institute (JASRI), Sayo 679-5198, Japan}

\author{Shunsuke Fukami}
\affiliation{Laboratory for Nanoelectronics and Spintronics, Research Institute of Electrical Communication, Tohoku University, Sendai 980-8577, Japan}
\affiliation{Center for Innovative Integrated Electronic Systems, Tohoku University, Sendai 980-0845, Japan}
\affiliation{Center for Spintronics Research Network, Tohoku University, Sendai 980-8577, Japan}
\affiliation{Center for Science and Innovation in Spintronics, Tohoku University, Sendai 980-8577, Japan}
\affiliation{WPI Advanced Institute for Materials Research, Tohoku University, Sendai 980-8577, Japan}

\author{Atsushi Fujimori}
\affiliation{Department of Physics, The University of Tokyo, Tokyo 113-0033, Japan}

\author{Hideo Ohno}
\affiliation{Laboratory for Nanoelectronics and Spintronics, Research Institute of Electrical Communication, Tohoku University, Sendai 980-8577, Japan}
\affiliation{Center for Innovative Integrated Electronic Systems, Tohoku University, Sendai 980-0845, Japan}
\affiliation{Center for Spintronics Research Network, Tohoku University, Sendai 980-8577, Japan}
\affiliation{Center for Science and Innovation in Spintronics, Tohoku University, Sendai 980-8577, Japan}
\affiliation{WPI Advanced Institute for Materials Research, Tohoku University, Sendai 980-8577, Japan}

\author{Masamitsu Hayashi}
\affiliation{Department of Physics, The University of Tokyo, Tokyo 113-0033, Japan}
\affiliation{National Institute for Materials Science, Tsukuba 305-0047, Japan}


\date{\today}

\begin{abstract}
We have studied the magnetic properties of multilayers composed of ferromagnetic metal Co and heavy metals with strong spin orbit coupling (Pt and Ir). 
Multilayers with symmetric (ABA stacking) and asymmetric (ABC stacking) structures are grown to study the effect of broken structural inversion symmetry.
We compare the perpendicular magnetic anisotropy (PMA) energy of symmetric Pt/Co/Pt, Ir/Co/Ir multilayers and asymmetric Pt/Co/Ir, Ir/Co/Pt multilayers. 
First, the interface contribution to the PMA is studied using the Co layer thickness dependence of the effective PMA energy.
Comparison of the interfacial PMA between the Ir/Co/Pt, Pt/Co/Ir asymmetric structures and Pt/Co/Pt, Ir/Co/Ir symmetric structures indicate that the broken structural inversion symmetry induced PMA is small compared to the overall interfacial PMA.
Second, we find the magnetic anisotropy field is significantly increased in multilayers when the ferromagnetic layers are antiferromagnetically coupled via interlayer exchange coupling (IEC).
Macrospin model calculations can qualitatively account for the relation between the anisotropy field and the IEC.
Among the structures studied, the IEC is the largest for the asymmetric Ir/Co/Pt multilayers: the exchange coupling field exceeds 3 T and consequently, the anisotropy field approaches 10 T. 
Third, comparing the asymmetric Ir/Co/Pt and Pt/Co/Ir structures, we find the IEC and, to some extent, the interface PMA are stronger for the former than the latter. 
X-ray magnetic circular dichroism (XMCD) studies suggest that the proximity induced magnetization in Pt is larger for the Ir/Co/Pt multilayers than the inverted structure (Pt/Co/Ir), which may partly account for the difference in the magnetic properties.
These results show the intricate relation between PMA, IEC and the proximity induced magnetization that can be exploited to design artificial structures with unique magnetic characteristics.
\end{abstract}


\maketitle

Perpendicular magnetic anisotropy (PMA) is one of the key parameters in developing modern spintronic devices including spin transfer torque (STT)-magnetic random access memory (MRAM)\cite{thomas2014jap}, spin orbit torque (SOT)-MRAM\cite{cubukcu2014apl,fukami2016nnano} and Racetrack memory\cite{parkin2008science,parkin2015nnano}. As STT or SOT is used to control the magnetization direction of the magnetic layer in these devices, it is common to use a few atomic layers thick magnetic layer to maximize the efficiency of current induced magnetization switching. For such ultrathin magnetic films, the magnetic easy axis lies along the film plane due to the strong shape anisotropy. A standard approach to changing the direction of the magnetic easy axis from the film plane to the film normal, which is beneficial for current controlled magnetization\cite{mangin2006nmat,ikeda2010nmat}, is to use the PMA that originates from interfaces. It is now well understood that certain combinations of materials give rise to a strong PMA\cite{carcia1985apl,denbroeder1988prl,denBroeder1991jmmm,daalderop1992prl,johnson1996rpp,nakajima1998prl,manchon2008jap,yakata2009jap,ikeda2010nmat,kubota2012jap}. In particular, the interface of a ferromagnetic metal (FM) with a heavy metal (HM) with strong spin orbit coupling is one of the prototypical systems to establish PMA in ultrathin magnetic films\cite{carcia1985apl,denbroeder1988prl,denBroeder1991jmmm,daalderop1992prl,johnson1996rpp,nakajima1998prl}.

The strength of PMA is directly related to the thermal stability of the magnetic bits in the MRAM/Racetrack technologies. It is often the case that one uses both the top and bottom interfaces of the FM layer to increase the overall PMA of the system\cite{sato2014apl}. For example, multilayers composed of repeats of FM/HM bilayers, such as Co/Pt multilayers, have been used as the reference layer of STT-MRAM. In addition, synthetic antiferromagnets with interlayer exchange coupling (IEC)\cite{parkin1990prl,parkin1991prl} are often used in the reference layer to reduce its stray field that may otherwise disturb the magnetization switching process of the information storage layer. Interestingly, it has been recently reported that antiferromagnets, including the synthetic antiferromagnets and ferrimagnets, can be used as the information storage layer in SOT-MRAM owing to the high efficiency to control magnetization via the SOT\cite{yang2015nnano,lau2016nnano,roschewsky2016apl,finley2016prap,mishra2017prl,blasing2018ncomm}.
It is thus of high importance to control both the PMA and the IEC in such systems.

The origin of PMA at the HM/FM and FM/oxide interfaces have been studied extensively\cite{bruno1989prb,daalderop1992prl,johnson1996rpp,nakamura2009prl,yang2011prb,dieny2017rmp}. In general, changes of the electron orbital occupation of atoms near the interface contribute to the PMA. It has been recently proposed that a spin-split band of the Rashba-type at the interface contributes to the emergence of PMA\cite{barnes2014scirep,kim2016prb}. The model predicts that structures with broken inversion symmetry, e.g. films with \textit{ABC} stacking, may lead to larger PMA than otherwise.
As such spin split states at the interface may also modify spin mixing conductance and the Dzyaloshinskii-Moriya interaction\cite{kim2013prl,manchon2015nmat}, interests in exploring structures with broken inversion symmetry are growing.

Here we present systematic studies on the magnetic properties of Pt/Co/Pt, Ir/Co/Ir symmetric multilayers and Pt/Co/Ir, Ir/Co/Pt asymmetric multilayers. We find a giant perpendicular magnetic anisotropy emerges in Ir/Co/Pt multilayers. The film stacking and thickness dependence of the PMA, IEC and the proximity induced magnetization are studied to identify the origin of the giant PMA.


Films were deposited, using RF magnetron sputtering, on silicon substrates coated with $\sim$100 nm thick silicon oxide. The film structure is sub./1 Ta/3 Ru/[$d_{\textrm{X}}$ X/$t$ Co/$d_{\textrm{Y}}$ Y]$_N$/2 MgO/1 Ta (thickness in units of nanometer), where X, Y = Ir, Pt. $N$ is the number of repeats of the unit structure denoted in the square brackets. $d_{\textrm{X}}$, $t$, and $d_{\textrm{Y}}$ are the nominal thickness of the X, Co, and Y layer, respectively. We refer to the films as \textit{symmetric} (\textit{asymmetric}) when X=Y (X$\neq$Y). The magnetic properties of the films were studied using vibrating sample magnetometry (VSM). Films with uniform thickness across the substrate were used for the VSM measurements. Since the applicable magnetic field of VSM is limited to $\sim$2 T, we use transport measurements when larger field is required. The transport measurements were carried out in physical property measurement system (PPMS) with maximum field of 14 T using patterned Hall bars. We define the $z$-axis as the direction parallel to the film plane normal. Current is passed along the $x$-axis and the Hall voltage is measured along the $y$-axis. Optical lithography and Ar ion etching were used to define the Hall bars and a standard liftoff process was used to pattern the electrodes (contact pads) made of 5 Ta/100 Au. Device patterning were performed on wedge films in which the thickness of X or Y layer is linearly varied across the substrate using a moving shutter during deposition of the film (the thickness of the rest of the layers are kept constant).

\begin{figure}[t]
	\includegraphics[scale=0.4]{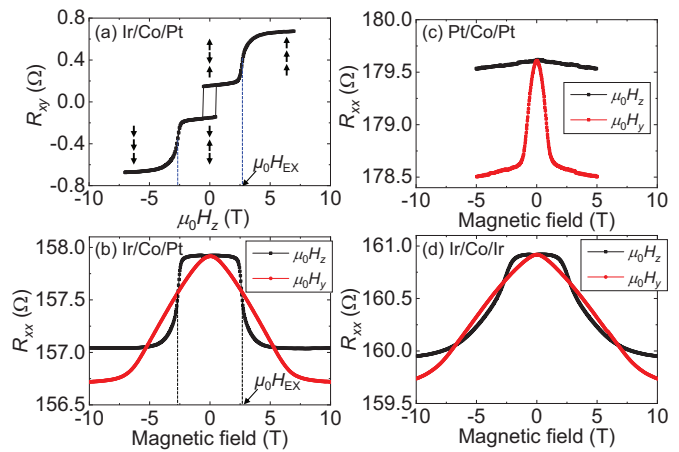}
	\caption{
	(a) The transverse resistance $R_{xy}$ plotted as a function of $\mu_0 H_z$ for a Hall bar made of 
[0.6 Ir/0.9 Co/0.13 Pt]$_3$. The arrows indicate the magnetization direction of the three Co layers. (b) $\mu_0 H_y$ (red circles) and $\mu_0 H_z$ (black squares) dependence of the longitudinal resistance $R_{xx}$. The device is the same with (a). Definition of $\mu_0 H_{\textrm{EX}}$ is schematically shown in (a,b). (c,d) $R_{xx}$ vs. $\mu_0 H_y$ (red circles) and $\mu_0 H_z$ (black squares) for Hall bars made of 
[0.6 Pt/0.9 Co/0.6 Pt]$_3$ (c) and 
[0.25 Ir/0.9 Co/0.25 Ir]$_3$ (d).\label{fig:loops}}
\end{figure}

The transverse resistance of a Hall bar  $R_{xy}$ is proportional to the $z$-component of the magnetization due to the anomalous Hall effect of the ferromagnetic layer:
\begin{equation}
\label{eq:Rxy}
R_{xy} = \Delta R_{\textrm{AHE}} \bar{m}_z,
\end{equation}
where $\Delta R_{\textrm{AHE}}$ is the anomalous Hall resistance and $\bar{m}_z$ is the average $z$-component of the magnetization of the FM layers in the system. Other contributions to $R_{xy}$ is neglected as the field sweep is limited along the $z$-axis for the $R_{xy}$ measurements.
Figure \ref{fig:loops}(a) shows $R_{xy}$ of a Hall bar made of $N=3$ asymmetric structure with an unit structure of [0.6 Ir/0.9 Co/0.13 Pt] plotted as a function of $\mu_0 H_z$.
As evident, the remanence value of $R_{xy}$ at zero $\mu_0 H_z$ is nearly one third of that of the largest $\mu_0 H_z$, suggesting that the three Co layers are coupled antiferromagnetically. The expected magnetic configuration of the three Co layers are illustrated using the arrows depicted in Fig.~\ref{fig:loops}(a). 
 
The longitudinal resistance $R_{xx}$ of the Hall bar is influenced by the current in-plane giant magnetoresistance (GMR) and the spin Hall magnetoresistance (SMR)\cite{nakayama2013prl,chen2013prb,kim2016prl}: 
\begin{equation}
\label{eq:Rxx}
R_{xx} = R_0 - \Delta R_{\textrm{GMR}} \cos\varphi - \Delta R_{\textrm{SMR}} \bar{m}_y^2,
\end{equation}
$R_0$ is the base resistance that does not depend on the magnetization, $\Delta R_{\textrm{GMR}}$ is the difference in $R_{xx}$ when the FM layers are in the anti-parallel and parallel states and $\Delta R_{\textrm{SMR}}$ is the spin  Hall magnetoresistance. $\varphi$ is the angle between the magnetization of the FM layers: when the FM layers are coupled ferromagnetically (anti-ferromagnetically), we define $\varphi=0$ ($\varphi=\pi$). $\bar{m}_y$ is the average $y$-component of the magnetization of FM layers in the system.
We have neglected contributions from the anisotropic magnetoresistance (AMR) here as the field sweep is limited within the $yz$ plane. 

The $\mu_0 H_z$ dependence of the longitudinal ($R_{xx}$) resistance of the same structure with Fig.~\ref{fig:loops}(a) ([0.6 Ir/0.9 Co/0.13 Pt]$_3$) is shown by the black squares in Fig.~\ref{fig:loops}(b). Resistance steps are found at $\mu_0 H_z$ $\sim$ $\pm 3$ T, which coincide with the switching field between the parallel and antiparallel states found in Fig.~\ref{fig:loops}(a). Thus the steps in $R_{xx}$ is caused by the GMR: the difference in $R_{xx}$ between the parallel ($\varphi=0$) and anti-parallel ($\varphi=\pi$) states is equal to $\Delta R_{\textrm{GMR}}$.
We define the transition field as the exchange coupling field $\mu_0 H_{\textrm{EX}}$ (see Fig.~\ref{fig:loops}(a) and \ref{fig:loops}(b) for an illustrative depiction of $\mu_0 H_{\textrm{EX}}$). 

The red circles in Fig.~\ref{fig:loops}(b) display $R_{xx}$ measured along against $\mu_0 H_y$. 
The magnetic configuration at zero field is the anti-parallel state (see the arrows shown in Fig.~\ref{fig:loops}(a)). As the magnitude of $\mu_0 H_y$ is increased, all moments will point along the field and the magnetic configuration becomes the parallel state. As evident, $R_{xx}$ decreases with increasing field and tends to saturate at $\mu_0 H_y \sim$7 T. This is the field at which all moments point along the $y$-axis: we define the saturation field (i.e. the magnetic anisotropy field) as $\mu_0 H_{\textrm{K}}$. Interestingly, $R_{xx}$ at $|\mu_0 H_y| >7$ T is not at the same level with that when $|\mu_0 H_z| >3$ T. The difference in $R_{xx}$ when large $H_y$ and $H_z$ are applied, caused by the SMR, corresponds to $\Delta R_{\textrm{SMR}}$ in Eq.~(\ref{eq:Rxx}).  
We verified that $\Delta R_{\textrm{SMR}}$ does not depend on the sign of IEC. $\Delta R_{\textrm{SMR}}$ is the smallest for the [Ir/Co/Ir] structures, increases when one of the Ir layer is replaced with Pt and is the largest for the [Pt/Co/Pt] structures. This is consistent with the trend one expects from the SMR since the spin Hall angle of Pt is significantly larger than that of Ir\cite{ishikuro2019prb}.

The magnetic field dependence of $R_{xx}$ for the $N=3$ symmetric structures with unit structures of [0.6 Pt/0.9 Co/0.6 Pt] and [0.25 Ir/0.9 Co/0.25 Ir] are shown in Figs. \ref{fig:loops}(c) and \ref{fig:loops}(d), respectively. For the former (X,Y=Pt), the easy axis loop ($R_{xx}$ vs. $\mu_0 H_z$) shows no step-like reduction of $R_{xx}$ with increasing $H_z$, suggesting that the three Co layers are not coupled antiferromagnetically.
From the hard axis field sweep (along $y$), we find $\mu_0 H_{\textrm{K}}$ $\sim$1 T, comparable to previous reports in similar structures\cite{bandiera2011iml,lee2014iml}. 
In contrast, for the Ir/Co/Ir symmetric structure, both the easy axis switching field and the hard axis saturation field are considerably larger. As reported previously, the IEC mediated by a thin Ir layer is one of the largest among the non-magnetic transition metals\cite{yakushiji2017apl}. Here, the total thickness of the Ir spacer is $\sim$ 0.5 nm, corresponding to the first antiferromagnetic coupling peak for Ir. The results in Fig.~\ref{fig:loops}(d) show that the anisotropy field $\mu_0 H_{\textrm{K}}$ is also enhanced, compared to the uncoupled Pt symmetric structure. Compared to the easy and hard axes loops of the asymmetric structure (Fig.~\ref{fig:loops}(b)), the loops of the Ir/Co/Ir symmetric structure are rounded and the easy axis switching/anisotropy fields are not well defined. 
In systems with $\mu_0 H_{\textrm{K}} < \mu_0 H_{\textrm{EX}}$, we find such loops indicating that the magnetic system becomes isotropic.


\begin{figure}[t]
	\includegraphics[scale=0.4]{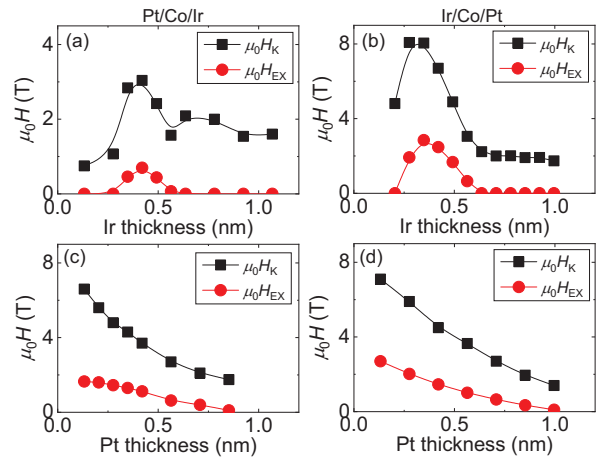}
	\caption{
	(a,b) The Ir layer thickness $d_{\textrm{Ir}}$ dependence of the anisotropy field $\mu_0 H_{\textrm{K}}$ (black squares) and the exchange coupling field $\mu_0 H_{\textrm{EX}}$ (red circles) for [0.6 Pt/0.9 Co/$d_{\textrm{Ir}}$ Ir]$_3$ (a) and [$d_{\textrm{Ir}}$ Ir/0.9 Co/0.6 Pt]$_3$ (b). (c,d) The Pt layer thickness $d_{\textrm{Pt}}$ dependence of the anisotropy field $\mu_0 H_{\textrm{K}}$ (black squares) and the exchange coupling field $\mu_0 H_{\textrm{EX}}$ (red circles) for [$d_{\textrm{Pt}}$ Pt/0.9 Co/0.5 Ir]$_3$ (c) and [0.6 Ir/0.9 Co/$d_{\textrm{Pt}}$ Pt]$_3$ (d). \label{fig:HkHex}}
\end{figure}

The Ir layer thickness dependence of the anisotropy field $\mu_0 H_{\textrm{K}}$ and the exchange coupling field $\mu_0 H_{\textrm{EX}}$ for the Pt/Co/Ir and Ir/Co/Pt asymmetric structures are shown in Figs. \ref{fig:HkHex}(a,b). The thickness of Co and Pt layers is fixed to $\sim$0.9 nm and $\sim$0.6 nm, respectively. $\mu_0 H_{\textrm{EX}}$ (red circles) shows a peak at $d_{\textrm{Ir}}\sim$0.4 nm. These results are consistent with previous reports in which a strong IEC was observed in Co/Pt multilayers coupled via a thin Ir layer\cite{itoh2003jmmm,yakushiji2017apl}.
We find a significantly larger $\mu_0 H_{\textrm{EX}}$ for the Ir/Co/Pt heterostructures compared to that of the Pt/Co/Ir heterostructures. 
Similar to the thickness dependence of $\mu_0 H_{\textrm{EX}}$, $\mu_0 H_{\textrm{K}}$ (black squares) also takes a maximum at $d_{\textrm{Ir}}$ $\sim$0.4 nm. Except for the hump-like structure at $d_{\textrm{Ir}}$ $\sim$0.7-0.9 nm for the Pt/Co/Ir heterostructure, $\mu_0 H_{\textrm{K}}$ is proportional to $\mu_0 H_{\textrm{EX}}$. For both structures, the large $d_{\textrm{Ir}}$ limit of $\mu_0 H_{\textrm{K}}$ takes a similar value, i.e. $\mu_0 H_{\textrm{K}}$ $\sim$2 T. However, significant difference in $\mu_0 H_{\textrm{K}}$ is found for films with thinner Ir. In particular, both $\mu_0 H_{\textrm{EX}}$ and $\mu_0 H_{\textrm{K}}$ are considerably larger for the Ir/Co/Pt heterostructures compared to its inverted counterpart when the Ir layer thickness is small ($d_{\textrm{Ir}}\sim$0.3-0.6 nm).

The Pt layer thickness dependence of $\mu_0 H_{\textrm{K}}$ and $\mu_0 H_{\textrm{EX}}$ are displayed in Figs. \ref{fig:HkHex}(c,d). The thickness of the Co layer is fixed to $\sim$0.9 nm and that of the Ir layer for Ir/Co/Pt (Pt/Co/Ir) is $\sim$0.6 nm ($\sim$0.5 nm). Within the thickness range of the Pt layer shown in Figs. \ref{fig:HkHex}(c,d), the Co layers are coupled antiferromagnetically. For both heterostructures, $\mu_0 H_{\textrm{K}}$ and $\mu_0 H_{\textrm{EX}}$ decay monotonically with increasing Pt layer thickness. These results show that it is mainly the Ir layer thickness that defines the sign of IEC but interestingly, combined with an Ir layer of proper thickness, a Pt layer of thickness up to $\sim$1 nm can mediate antiferromagnetic IEC even though Pt alone exhibits weak ferromagnetic IEC without oscillation\cite{parkin1991prl,liu2004prb,knepper2005prb,zhao2008jap,bandiera2012apl}.

\begin{figure}[t]
	\includegraphics[scale=0.4]{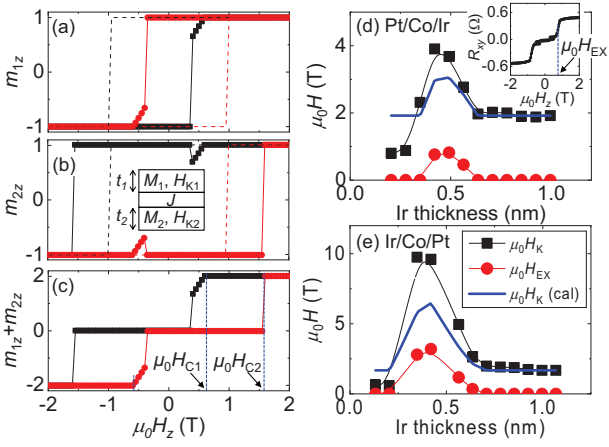}
	\caption{
	(a-c) Numerically calculated $z$-component of the magnetization of layer 1 (a), layer 2 (b) and the sum of the two (c). The field sweep direction is from positive to negative field for the black squares and from negative to positive field for the red circles. The inset of (b) schematically shows the system used for the calculations; i.e. two FM layers coupled via the IEC with strength $J$. $\mu_0 H_{\textrm{C1}}$ and $\mu_0 H_{\textrm{C2}}$ are schematically shown in the inset of (c). Parameters used in the numerical calculations (solid lines): $M_1$: 1000 kA/m, $M_2$: 1010 kA/m, $\mu_0 H_{\textrm{K1}}$: 1 T, $\mu_0 H_{\textrm{K2}}$: 1 T, $\alpha_1$: 1, $\alpha_2$: 1, $t_1$: 1 nm, $t_2$: 1 nm, $J$: -0.8 mJ/m$^2$. The dashed lines in (a) and (b) show the case for $J=0$. (d,e) The Ir layer thickness $d_{\textrm{Ir}}$ dependence of the anisotropy field $\mu_0 H_{\textrm{K}}$ (black squares) and the exchange coupling field $\mu_0 H_{\textrm{EX}}$ (red circles) for [0.6 Pt/0.9 Co/$d_{\textrm{Ir}}$ Ir]$_{12}$ (d) and [$d_{\textrm{Ir}}$ Ir/0.9 Co/0.6 Pt]$_{12}$ (e). The blue solid line shows calculated $\mu_0 H_{\textrm{K}}$ using Eq.~(\ref{eq:HK2N}). Inset of (d): $R_{xy}$ vs. $\mu_0 H_z$ for a Hall bar made of [0.6 Pt/0.9 Co/0.42 Ir]$_{12}$. \label{fig:Multilayers}}
\end{figure}

To study the correlation between the PMA (anisotropy field) and the IEC, we use a macrospin model. We start from a system that consists of two ferromagnetic layers ($i=1,2$) separated by a non-magnetic spacer layer: see the inset of Fig.~\ref{fig:Multilayers}(b). In accordance with the experiments, we  set the magnetic easy axis of the two layers along the $z$ axis. The magnetic energy density of the system reads
\begin{equation}
\begin{aligned}
\frac{E}{A}&=-\mu_0 \bm{M}_1 \cdot \bm{H} t_1 - \mu_0 \bm{M}_2 \cdot \bm{H} t_2 - \frac{1}{2} \mu_0 M_1 H_{\textrm{K}1} t_1 \cos^2\theta_1\\
&- \frac{1}{2} \mu_0 M_2 H_{\textrm{K}2} t_2 \cos^2\theta_2 - J \bm{m}_1 \cdot \bm{m}_2,
\label{eq:energy}
\end{aligned}
\end{equation}
where $\bm{M}_i$ is the magnetization vector of layer $i$, $\bm{m}_i$ is its unit vector and $\theta_i$ is the polar angle of $\bm{m}_i$ with respect to the $z$ axis. $M_i$, $H_{\textrm{K}i}$ and $t_i$ are the saturation magnetization, uniaxial anisotropy field and the thickness, respectively, of layer $i$. $J$ is the interlayer exchange coupling constant between layers 1 and 2. The sign of $J$ is defined positive (negative) if the magnetization of the two magnetic layers  exhibit ferromagnetic (anti-ferromagnetic) coupling. $A$ is the area of the interface and $\bm{H}$ is the applied magnetic field.

The Landau-Lifshitz-Gilbert equation,
\begin{equation}
\frac{\partial \bm{m}}{\partial t} = - \gamma \bm{m} \times \frac{\partial E}{\partial \bm{m}} + \alpha \frac{\partial \bm{m}}{\partial t},
\label{eq:llg}
\end{equation}
is numerically solved with the parameters defined in the caption of Fig.~\ref{fig:Multilayers} to calculate the magnetization hysteresis loops of the two FM layers 1 and 2 with a negative $J$. The calculated $z$-component of magnetization of the two layers, $m_{1z}$ and $m_{2z}$, and the sum, $m_{1z} + m_{2z}$, are plotted as a function of $\mu_0 H_z$ in Figs. \ref{fig:Multilayers}(a-c). Although the magnetic parameters of the two FM layers are set to similar values, the magnetization switching fields are different due to the IEC.
Using energy minimization and looking for the condition at which one solution becomes more stable than the others, the switching field from the parallel to antiparallel configuration is expressed as \cite{knepper2005prb}
\begin{equation}
\begin{aligned}
H_{\textrm{C1}} &= -H_{\textrm{K}0} + (H_{\textrm{J}1} + H_{\textrm{J}2}),
\label{eq:H1}
\end{aligned}
\end{equation}
where $H_{\textrm{J}i} \equiv -\frac{J}{\mu_0 M_i t_i}$ ($i=1,2$). Similarly, the field at which transition from the anti-parallel to parallel configuration takes place is given by
\begin{equation}
\begin{aligned}
H_{\textrm{C2}} = &- \frac{H_{\textrm{J}1} - H_{\textrm{J}2}}{2} + \Big\{ H_{\textrm{K}0}^2 + H_{\textrm{K}0} \big(H_{\textrm{J}1} + H_{\textrm{J}2} \big)\\
&\ \ \ \ \ + \frac{(M_1-M_2)^2 H_{\textrm{J}1} H_{\textrm{J}2}}{4 M_1 M_2} \Big \}^{1/2}.
\label{eq:H2}
\end{aligned}
\end{equation}
In deriving Eqs. (\ref{eq:H1}) and (\ref{eq:H2}), we have assumed $H_{\textrm{K}1} = H_{\textrm{K}2} \equiv H_{\textrm{K}0}$ and $t_1=t_2 \equiv t$.
This is typically the case since, in experiments, the magnetic layer consists of the same material with the same thickness. Note that the switching field shown above only applies to system with negative $J$.
$\mu_0 H_{\textrm{C1}}$ and $\mu_0 H_{\textrm{C2}}$ are schematically depicted in Fig.~\ref{fig:Multilayers}(c).

The apparent hard axis magnetic anisotropy field $\mu_0 H_{\textrm{K}}$ is defined as the in-plane magnetic field needed to cause the magnetization to point along the film plane. Without the interlayer exchange coupling, $H_{\textrm{K}}$ is equal to $H_{\textrm{K}0}$. With the IEC, $H_{\textrm{K}}$ takes the form:
\begin{equation}
\begin{aligned}
H_{\textrm{K}} = H_{\textrm{K}0} + H_{\textrm{J}1} + H_{\textrm{J}2}.
\label{eq:HK}
\end{aligned}
\end{equation}

From hereon, we also assume $M_1 \approx M_2 \equiv M$ and define $H_{\textrm{J}} \equiv -\frac{J}{\mu_0 M t}$. 
Substituting these conditions into Eqs. (\ref{eq:H1}) and (\ref{eq:H2}), the exchange field defined below reads
\begin{equation}
\begin{aligned}
H_{\textrm{EX}} &\equiv  \frac{1}{2} \big( H_{\textrm{C2}} + H_{\textrm{C1}} \big) \\
&\approx \frac{1}{2} \big(\sqrt{H_{\textrm{K}0}^2 + 2 H_{\textrm{K}0} H_{\textrm{J}}} - (H_{\textrm{K}0} - 2 H_{\textrm{J}}) \big).
\label{eq:Hex}
\end{aligned}
\end{equation}
Similarly, the apparent hard axis magnetic anisotropy field $H_{\textrm{K}}$ becomes
\begin{equation}
\begin{aligned}
H_{\textrm{K}} &\approx H_{\textrm{K}0} + 2 H_{\textrm{J}}.
\label{eq:HK2}
\end{aligned}
\end{equation}
As evident, $H_{\textrm{EX}} \rightarrow 0$ and $H_{\textrm{K}} \rightarrow H_{\textrm{K}0}$ when $J \rightarrow 0$.
Equation (\ref{eq:HK2}) also shows that, with the IEC, $H_{\textrm{K}}$ increases linearly with $J$. 
This can be understood as following: on applying an in-plane external field, the magnetization of each layer deviates from the initial perpendicular collinear antiparallel state to adopt a non-collinear scissored arrangement with finite in-plane component along the field direction. Similar to the uniaxial magnetic anisotropy, IEC favoring antiparallel arrangement acts as an additional restoring force against the external field, leading to an enhancement of the apparent $H_{\textrm{K}}$.
Note that spin-flop type transitions\cite{engel2005ieee} are not observed under application of hard axis field here.  

We may rewrite Eqs. (\ref{eq:Hex}) and (\ref{eq:HK2}) to obtain the following relation between $H_\textrm{K}$ and $H_{\textrm{EX}}$:
\begin{equation}
\begin{aligned}
H_{\textrm{K}} \approx H_{\textrm{K}0} + \frac{1}{2} \big[ 3 H_{\textrm{K0}} + 4 H_{\textrm{EX}} - \sqrt{9 H_{\textrm{K0}}^2 + 8 H_{\textrm{K0}} H_{\textrm{EX}}} \big].
\label{eq:HK2N}
\end{aligned}
\end{equation}
The model above describes IEC in a trilayer system with two FM layers separated by a non-magnetic spacer layer.
In FM/NM multilayers (i.e. multiple repeats of the FM/NM unit structure), the IEC becomes twice as large compared to that of the trilayer system since the number of exchange coupled interfaces almost doubles (we ignore contribution from the edge FM layers).
Thus the relation between $H_{\textrm{K}}$ and $H_{\textrm{EX}}$, as denoted in Eq.~(\ref{eq:HK2N}), holds for the multilayer system as well.
We may therefore test this relation using heterostructures with increased number of $N$.

\begin{figure}[t]
	\includegraphics[scale=0.5]{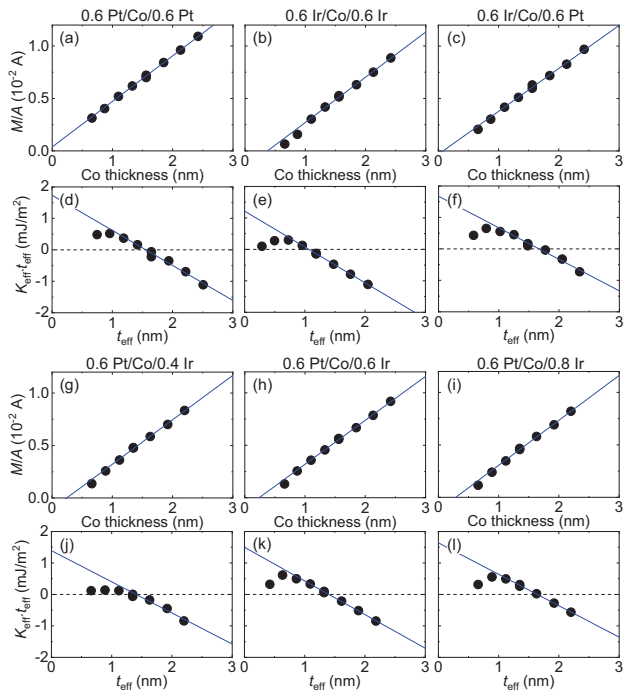}
	\caption{
	(a-c) Co layer thickness $t$ dependence of magnetization per unit area $M/A$ (a-c, g-i) and the effective FM layer thickness $t_{\textrm{eff}}$ dependence of product of the effective magnetic anisotropy energy density $K_{\textrm{eff}}$ and $t_{\textrm{eff}}$ (d-f, j-l). The film structure is [0.6 Pt/$t$ Co/0.6 Pt]$_3$ (a,d), [0.6 Ir/$t$ Co/0.6 Ir]$_3$ (b,e), [0.6 Ir/$t$ Co/0.6 Pt]$_3$ (c,f), [0.6 Pt/$t$ Co/$d_{\textrm{Ir}}$ Ir]$_3$ (g-l). The Ir layer thickness $d_{\textrm{Ir}}$ is 0.4 nm (g,j), 0.6 nm (h.k) and 0.8 nm (i,l). The blue solid lines represent linear fit to the data of appropriate range.   \label{fig:Ki}}
\end{figure}

Black squares and red circles in Figs.~\ref{fig:Multilayers}(d,e) show the Ir layer thickness dependence of experimentally obtained $\mu_0 H_{\textrm{K}}$ and $\mu_0 H_{\textrm{EX}}$, respectively, for the Pt/Co/Ir and Ir/Co/Pt asymmetric structures with $N=12$. Compared to the heterostructures with $N=3$ (Figs.~\ref{fig:HkHex}(a) and \ref{fig:HkHex}(b)), $\mu_0 H_{\textrm{K}}$ and $\mu_0 H_{\textrm{EX}}$ are both larger. Such trend may be attributed to the enhanced fcc(111) texture as the overall film thickness becomes larger. The calculated $\mu_0 H_{\textrm{K}}$ using Eq.~(\ref{eq:HK2N}) is shown by the blue solid line in Figs. \ref{fig:Multilayers}(d) and \ref{fig:Multilayers}(e).
In the calculation we assume a constant $\mu_0 H_{\textrm{K}0}$ that is taken from the thick Ir limit of $\mu_0 H_{\textrm{K}}$ found experimentally. The calculated results are in relatively good agreement with those of the experimental results. However, we find that the calculation tends to underestimate $\mu_0 H_{\textrm{K}}$ found in the experiments when the antiferromagnetic IEC is strong.

These results show that the large $\mu_0 H_{\textrm{K}}$ of the asymmetric structures is predominantly due to the existence of IEC.
Next we evaluate the interface contribution to $\mu_0 H_{\textrm{K}}$ by studying the Co layer thickness ($t$) dependence of the magnetic properties.
Magnetization hysteresis loops with field swept along the magnetic easy and hard axes are measured using VSM.
Figures~\ref{fig:Ki}(a-c) and \ref{fig:Ki}(g-i) show the the measured magnetic moment per unit area $M/A$ of the symmetric and asymmetric structures plotted as a function of $t$. The thickness of the X and Y layers in the heterostructures is denoted in each panel and the number of repeat $N$ of the unit structure is 3. As evident, $M/A$ increases linearly with $t$: we fit the data with a linear function to find the saturation magnetization $M_{\textrm{S}}$ and the magnetic dead layer thickness $t_{\textrm{D}}$ from the slope and the $x$-axis intercept of the fitted linear function, respectively. 

The effective magnetic anisotropy energy $K_{\textrm{eff}}$ of the film is obtained by calculating the areal difference between the out-of-plane and in-plane magnetization hysteresis loops. Note that $K_{\textrm{eff}}$ estimated using the areal difference excludes contribution from the interlayer exchange coupling. The product $K_{\textrm{eff}} t_{\textrm{eff}}$ is plotted as a function of $t_{\textrm{eff}} \equiv t - t_{\textrm{D}}$ in Figs. \ref{fig:Ki}(d-f) and (j-l). (See the Appendix for comparison of all structures with $d_{\textrm{X}} = d_{\textrm{Y}} = 0.6$ nm.) As with the case of other systems\cite{johnson1995jmmm,ikeda2010nmat,sinha2013apl}, $K_{\textrm{eff}} t_{\textrm{eff}}$ increases with decreasing $t_{\textrm{eff}}$ before it drops when the magnetic layer becomes thin and/or when strain effects take place\cite{johnson1995jmmm,gowtham2016prb,lau2017apl}. We therefore fit the data in the appropriate thickness range with a linear function. The slope and the $y$-axis intercept of the fitted linear function provide information on the bulk contribution $K_{\textrm{B}} - \frac {1}{2} \mu_0 M_{\textrm{S}}^2$ and interface contribution $K_{\textrm{I}}$ to the magnetic anisotropy energy. The parameters obtained from the fittings are summarized in Table~\ref{table:Ki}.

\begin{table}[b]
\caption{Saturation magnetization $M_{\textrm{S}}$, magnetic dead layer thickness $t_{\textrm{D}}$, bulk $K_{\textrm{B}}$ and interface $K_{\textrm{I}}$ contributions to the magnetic anisotropy energy of the symmetric and asymmetric structures. \label{table:Ki}}
\begin{ruledtabular}
	\begin{tabular}{ccccc}
	Samples & $M_{\textrm{S}}$ & $t_{\textrm{D}}$ & $K_{\textrm{B}}$ & $K_{\textrm{I}}$\\
	     & kA/m & nm & 10$^5$ J/m$^3$ & mJ/m$^2$\\
	\hline
	0.6 Pt/Co/0.6 Pt &	1450	&	-0.1	&	2.0	&	1.7 \\
	0.6 Ir/Co/0.6 Ir	&	1440	&	0.4	&	1.6	&	1.2 \\
	0.6 Ir/Co/0.6 Pt	&	1370	&	0.1	&	1.7	&	1.7 \\
	0.6 Pt/Co/0.6 Ir	&	1400	&	0.2	&	1.6	&	1.5 \\
	\hline
	0.6 Pt/Co/0.4 Ir	&	1410	&	0.2	&	2.7	&	1.4 \\
	0.6 Pt/Co/0.8 Ir	&	1430	&	0.3	&	2.8	&	1.6 \\
	\end{tabular}
\end{ruledtabular}
\end{table}

Table~\ref{table:Ki} shows that the average saturation magnetization $M_{\textrm{S}}$ is close to that of bulk Co for all structures. The magnetic dead layer thickness is non-zero when Co is placed next to Ir. The dead layer thickness is the largest for the Ir/Co/Ir symmetric structure and $t_\textrm{D}$ seems to be larger when Ir is placed on top of Co compared to the case when it sits below Co.
A negative $t_{\textrm{D}}$ is found for the Pt/Co/Pt symmetric structures, which is due to proximity induced magnetization of Pt.
The interface contribution ($K_{\textrm{I}}$) to the magnetic anisotropy energy is all positive and is the largest for the Pt/Co/Pt symmetric structure.
In contrast, the bulk contribution ($K_{\textrm{B}}$) is found to be small compared to the overall magnetic anisotropy energy.

Note that the $K_{\textrm{I}}$ values listed in Table~\ref{table:Ki} are not sufficient to unambiguously determine the contribution from each interface.
However, we may check the presence of PMA that originates from electronic states in broken structural inversion symmetry\cite{barnes2014scirep,kim2016prb,pradipto2019prb}. To do so, we compare the average $K_{\textrm{I}}$ of the symmetric and asymmetric structures using the values from Table~\ref{table:Ki}. The average $K_{\textrm{I}}$ of Pt/Co/Pt and Ir/Co/Ir structures is $\sim$1.5 mJ/m$^2$, and the average $K_{\textrm{I}}$ of Ir/Co/Pt and Pt/Co/Ir structures is $\sim$1.6 mJ/m$^2$. Although the average $K_{\textrm{I}}$ is $\sim$0.1 mJ/m$^2$ larger for the asymmetric structure, the difference is small compared to the overall PMA. From these results, it is difficult to conclude, in part due to the uncertainties in extracting $K_\textrm{I}$ by linear extrapolation, whether the broken structural inversion symmetry induced PMA is present in the heterostructures studied here.

As shown in Fig.~\ref{fig:Multilayers}(d) and \ref{fig:Multilayers}(e), the calculated $\mu_0 H_\textrm{K}$ based on Eq.~(\ref{eq:HK2N}) underestimates the anisotropy field for films with thinner Ir layer. Table~\ref{table:Ki} shows that for the Pt/Co/Ir multilayers with different Ir thicknesses ($d_{\textrm{Ir}}$ $\sim$0.4, 0.6, 0.8 nm), $K_{\textrm{I}}$ is close $\sim$1.5 mJ/m$^2$ and is not necessarily larger for the thinner Ir films. These results justify the assumption used in the calculations of $\mu_0 H_{\textrm{K}}$ that $\mu_0 H_{\textrm{K}0}$ is constant as a function Ir thickness. Thus the difference between the calculated and experimentally obtained $\mu_0 H_{\textrm{K}}$ cannot be accounted for with the current parameters. We therefore infer that either the macrospin model (i.e. Eq. (10)) is not sufficient to describe the relation between the anisotropy field $\mu_0 H_{\textrm{K}}$ and the exchange coupling field $\mu_0 H_{\textrm{EX}}$, or other sources of PMA exist for the thin Ir asymmetric structures. The former may be associated with the magnetization switching process in which the calculations assume a single domain Stoner-Wohlfarth type switching whereas in the experiments, the reversal process may involve domain wall nucleation and propagation. The non-collinear antiferromagnetic states that appear upon application of magnetic field may also contribute to the discrepancy.

\begin{figure}[t]
	\includegraphics[scale=0.4]{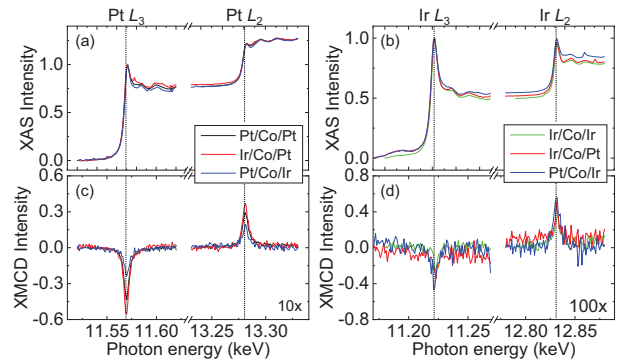}
	\caption{
	(a-d) X-ray absorption spectra (a,b) and X-ray magnetic circular dichroism spectra (c,d) at the Pt $L_3$, $L_2$ edges (a,c) and at the Ir $L_3$, $L_2$ edges (b,d). The film structure is [0.6 Pt/0.9 Co/0.6 Pt]$_1$ (black), [0.6 Ir/0.9 Co/0.6 Pt]$_1$ (red), [0.6 Pt/0.9 Co/0.6 Ir]$_1$ (blue), and [0.6 Ir/0.9 Co/0.6 Ir]$_1$ (green). The XMCD spectra are multiplied by a factor of 10 (c) and 100 (d). \label{fig:MCD}}
\end{figure}

\begin{table}[b]
\caption{Spin $m_{\textrm{s}}$ and orbital $m_{\textrm{o}}$ magnetic moments of Pt and Ir in the symmetric and asymmetric structures. The unit of the magnetic moments is $\mu_{\textrm{B}}$/hole. The error due to uncertainty of the signal analyses (background subtraction from the spectra) is $\sim$10\% for the Pt $m_{\textrm{s}}$ (the measurement error is smaller). For the Pt $m_{\textrm{o}}$ and the Ir moments, the error bar is of the same order of magnitude with the measured moment size. \label{table:MCD}}
\begin{ruledtabular}
	\begin{tabular}{ccccc}
	Samples & Pt $m_{\textrm{s}}$ & Pt $m_{\textrm{o}}$ & Ir $m_{\textrm{s}}$ & Ir $m_{\textrm{o}}$\\
	\hline
	Pt/Co/Pt	&	0.23	&	0.03	&	n/a	&	n/a\\
	Ir/Co/Ir	&	n/a	&	n/a	&	0.01	&	-0.00 \\
	Ir/Co/Pt	&	0.30	&	0.04	&	0.01	&	-0.00 \\
	Pt/Co/Ir	&	0.13	&	0.01	&	0.03	&	-0.00 \\
	\end{tabular}
\end{ruledtabular}
\end{table}

Interestingly, we find that the IEC and the interface PMA ($K_{\textrm{I}}$)
are larger for the Ir/Co/Pt asymmetric structures compared to its inverted structure, Pt/Co/Ir. To identify the origin of the difference in IEC and PMA for the two asymmetric structures (Ir/Co/Pt and Pt/Co/Ir), we have studied the spin and orbital magnetic moments of Pt and Ir in the heterostructures using X-ray magnetic circular dichroism (XMCD) spectroscopy. The thickness of the X and Y layers are fixed to $\sim$0.6 nm and the Co layer thickness is $\sim$0.9 nm. The number of repeats of the unit structure $N$ is fixed to 1. (Similar results are found in the Pt/Co/Ir heterostructures with $N=1$ and $N=2$.) Magnetic field ($\sim$2 T) perpendicular to the film plane is applied during the measurements. The X-ray absorption spectra (XAS) of Pt and Ir at the $L_{3}$ and $L_{2}$ edges are shown in Figs.~\ref{fig:MCD}(a) and \ref{fig:MCD}(b), respectively. We find similar XAS for all heterostructures. In contrast, the XMCD spectra show clear dependence on the film stacking. We estimate the spin $m_{\textrm{s}}$ and orbital $m_{\textrm{o}}$ magnetic moments of Pt and Ir using the XMCD spectra and the magneto-optical sum rules\cite{chen1995prl}. The values are summarized in Table~\ref{table:MCD}. Since we use the partial fluorescence yield mode to collect the signal, which gives negligible depth dependence of the XMCD signal strength as compared to the electron yield mode, the obtained magnetic moments are not influenced by the element's distance to the surface for the structures studied here and they reflect the thickness averaged values.

As evident from Table~\ref{table:MCD}, we find non-negligible difference in the Pt $m_{\textrm{s}}$ for the heterostructures containing Pt: $m_{\textrm{s}}$ is the largest for the Ir/Co/Pt heterostructure and is the smallest for the Pt/Co/Ir heterostructure. $m_{\textrm{s}}$ for the Pt/Co/Pt symmetric structure is close to the average value of the Ir/Co/Pt and Pt/Co/Ir asymmetric structures. These results thus show that the Pt $m_{\textrm{s}}$ is larger when it is deposited on Co compared to the case when it is placed below. Similar to $m_{\textrm{s}}$, the Pt $m_{\textrm{o}}$ is also larger when Pt is placed on top of Co. For Ir, we find order of magnitude smaller spin and orbital magnetic moment with similarly larger moments at the top interface (the values are close to the detection limit of XMCD). 
Note that the magnetic dead layer thickness of the Ir/Co/Ir structures is non-zero ($t_D\sim$0.4 nm, see Table~\ref{table:Ki}). The coexistence of strong IEC, moderate PMA and relatively thick magnetic dead layer in the symmetric Ir/Co/Ir multilayers suggests that interface with a finite magnetically dead layer does not necessarily imply the absence of interfacial PMA, IEC and vice versa\cite{kolesnikov2016jpd}.

Theoretically, it has been suggested that a spin-split band of the Rashba-type at the interface may lead to enhanced induced orbital moment of the heavy metal layer in contact with ferromagnetic layers\cite{grytsyuk2016prb}. The larger Pt $m_{\textrm{o}}$ at the top Co/Pt interface can therefore be reflecting the presence of such spin-split band at the interface, which may also enhance the PMA via the strong spin orbit coupling of such states. This may account for the larger $K_{\textrm{I}}$ found for the Ir/Co/Pt heterostructures compared to the Pt/Co/Ir heterostructures. It is unclear whether the IEC can be influenced by such spin-split states at the interface. As IEC has been generally known to originate from exchange of spin current across the non-magnetic spacer layer\cite{bruno1995prb,stiles1999jmmm}, strong spin orbit coupling of the spacer layer has been considered to be detrimental to the IEC. On the other hand, a larger induced moment in Pt may be beneficial for mediating the strong antiferromagnetic IEC of the Ir spacer, thus explaining the larger $\mu_0 H_\textrm{EX}$ and its slower decay with increasing Pt thickness, observed for Ir/Co/Pt multilayers compared to the inverted Pt/Co/Ir multilayers. We infer that the spin-split states at the top interface may contribute to both the large PMA and IEC in the Ir/Co/Pt multilayers.

Finally, it is worth noting that seed layer of different surface energy may change the growth mode, strain and degree of intermixing of the resulting sputtered films. Such stacking order dependence of the structural properties can influence magnetic properties related to interface states. For example, previous studies have shown that the interfacial Dzyaloshinskii-Moriya interaction (DMI) in sputtered Pt/Co/Pt symmetric heterostructures is not zero \cite{hrabec2014prb,je2013prb,wells2017prb}, suggesting that the DMI at the lower Pt/Co and the upper Co/Pt interfaces are not identical (the sign of DMI at the lower and upper interface is opposite). Asymmetry in the proximity induced magnetization was also found in Pd/Co/Pd symmetric structures\cite{kim2016scirep}: induced moment of Pd at the upper interface is larger than that of the bottom interface albeit little difference in the roughness at the two interfaces. Note that the different magnetic properties of Ir/Co/Pt and Pt/Co/Ir multilayers do not alter with increasing repetition number $N$; see Figs.~\ref{fig:HkHex}((a,b) and \ref{fig:Multilayers}(d,e). Since any difference in the texture or strain will be mitigated as the film thickness increases by increasing $N$, we infer it is not the texture or the strain that causes the difference. We consider growth-related intermixing at the interface and consequently different electronic structure of the interface states play a certain role for the difference in PMA and IEC of the Ir/Co/Pt and Pt/Co/Ir asymmetric structures. 


In summary, we have studied the perpendicular magnetic anisotropy (PMA) and the interlayer exchange coupling (IEC) in symmetric and asymmetric heterostructures. 
We find a giant perpendicular magnetic anisotropy field ($\mu_0 H_\textrm{K}$) emerges in Ir/Co/Pt multilayers: $\mu_0 H_\textrm{K}$ approaches 10 T for an optimized structure.
To identify its origin, film stacking and thickness dependence of the PMA, IEC and the proximity induced magnetization are studied.
Direct comparison of the interfacial PMA between the Ir/Co/Pt, Pt/Co/Ir asymmetric structures and Pt/Co/Pt, Ir/Co/Ir symmetric structures indicate that the broken structural inversion symmetry induced PMA is rather small.
We find that $\mu_0 H_\textrm{K}$ is significantly increased when the ferromagnetic layers are antiferromagnetically coupled via interlayer exchange coupling (IEC).
The Ir layer thickness dependence of $\mu_0 H_\textrm{K}$ and the exchange coupling field $\mu_0 H_\textrm{EX}$ show a direct correlation, which can be qualitatively accounted for using macrospin model calculations. 
The IEC and consequently $\mu_0 H_\textrm{K}$ are found to be significantly larger for the Ir/Co/Pt multilayers compared to the inverted Pt/Co/Ir multilayers. 
X-ray magnetic circular dichroism (XMCD) spectroscopy studies indicate that the proximity induced magnetization of Pt is also larger for the Ir/Co/Pt structures than that of the inverted Pt/Co/Ir structures. 
These results thus show the correlation between PMA, IEC and the proximity induced magnetization that may originate from the spin split states at the interface.
Such characteristics of artificial heterostructures can be exploited to design structures with strong PMA and IEC, potentially useful for magnetic memory technologies.

\section*{Appendix}
\subsection{Comparison of $K_\textrm{eff}$}
\begin{figure}[t]
	\includegraphics[scale=0.9]{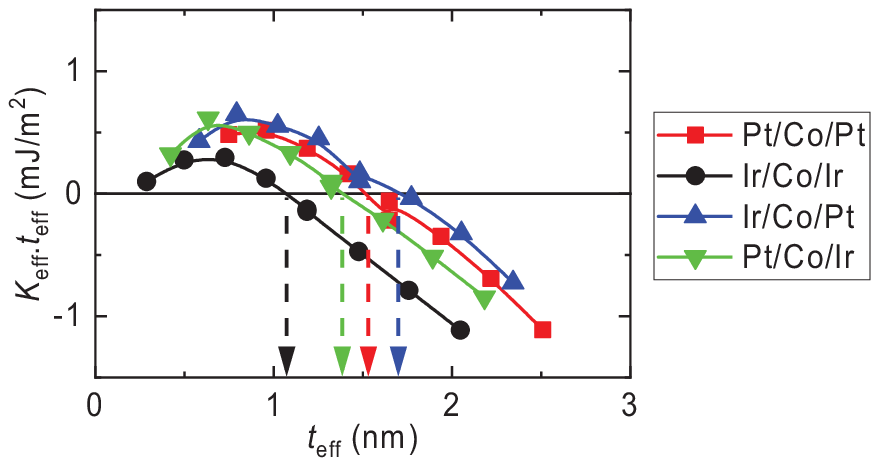}
	\caption{
	The effective Co layer thickness $t_{\textrm{eff}}$ dependence of the effective magnetic anisotropy energy density $K_{\textrm{eff}} t_{\textrm{eff}}$. The film structure is [0.6 Pt/$t$ Co/0.6 Pt]$_3$ (red squares), [0.6 Ir/$t$ Co/0.6 Ir]$_3$ (black circles), [0.6 Ir/$t$ Co/0.6 Pt]$_3$ (blue up triangles), [0.6 Pt/$t$ Co/0.6 Ir]$_3$ (green down triangles). The arrows indicate the crossover thicknesses where $K_\textrm{eff} t_{\textrm{eff}}$ changes sign.  
	\label{fig:Keff_all}}
\end{figure}

To provide a direct visual comparison of the PMA energy density among the multilayers with different stacking, we overlay in Fig.~\ref{fig:Keff_all} the $t_{\textrm{eff}}$ dependence of $K_{\textrm{eff}} t_{\textrm{eff}}$ for multilayers with $d_\textrm{X} = d_\textrm{Y} \sim 0.6$ nm. 
For each data set, we extract the critical effective Co thickness $t_\textrm{eff}$ at which $K_{\textrm{eff}} t_{\textrm{eff}}$ changes sign. These crossover thicknesses are indicated by vertical arrows of appropriate colors in Fig.~\ref{fig:Keff_all}. As evident, the asymmetric Ir/Co/Pt structures show the largest PMA energy density, with a maximum $K_{\textrm{eff}} t_{\textrm{eff}}$ of $\sim 0.65 $ mJ/m$^2$ and a crossover thickness of $\sim 1.7$ nm. (The corresponding $K_{\textrm{eff}}$ is $\sim 8.2 \times 10^5$ J/m$^3$ at $t_\textrm{eff} \sim 0.8$ nm (or $t \sim 0.9$ nm)). 

\begin{acknowledgments}
Acknowledgments: We thank K. Goto, Y. Nonaka, K. Ikeda, M. Mizumaki for technical support and S. Mitani for helpful discussion. This work was partly supported by JSPS Grant-in-Aid for Specially Promoted Research (15H05702), Scientific Research (16H03853), Casio Foundation, and the Center of Spintronics Research Network of Japan. The XMCD experiment was performed at BL39XU of SPring-8 with the approval of the Japan Synchrotron Radiation Research Institute (JASRI) (Proposal Nos. 2017A1048 and 2018A1058). Y.-C.L. is supported by JSPS International Fellowship for Research in Japan (Grant No. JP17F17064). Z.C. acknowledges financial support from Materials Education program for the future leaders in Research, Industry, and Technology (MERIT).
\end{acknowledgments}

\bibliography{reference_101919}

\begin{thebibliography}{60}%
\makeatletter
\providecommand \@ifxundefined [1]{%
 \@ifx{#1\undefined}
}%
\providecommand \@ifnum [1]{%
 \ifnum #1\expandafter \@firstoftwo
 \else \expandafter \@secondoftwo
 \fi
}%
\providecommand \@ifx [1]{%
 \ifx #1\expandafter \@firstoftwo
 \else \expandafter \@secondoftwo
 \fi
}%
\providecommand \natexlab [1]{#1}%
\providecommand \enquote  [1]{``#1''}%
\providecommand \bibnamefont  [1]{#1}%
\providecommand \bibfnamefont [1]{#1}%
\providecommand \citenamefont [1]{#1}%
\providecommand \href@noop [0]{\@secondoftwo}%
\providecommand \href [0]{\begingroup \@sanitize@url \@href}%
\providecommand \@href[1]{\@@startlink{#1}\@@href}%
\providecommand \@@href[1]{\endgroup#1\@@endlink}%
\providecommand \@sanitize@url [0]{\catcode `\\12\catcode `\$12\catcode
  `\&12\catcode `\#12\catcode `\^12\catcode `\_12\catcode `\%12\relax}%
\providecommand \@@startlink[1]{}%
\providecommand \@@endlink[0]{}%
\providecommand \url  [0]{\begingroup\@sanitize@url \@url }%
\providecommand \@url [1]{\endgroup\@href {#1}{\urlprefix }}%
\providecommand \urlprefix  [0]{URL }%
\providecommand \Eprint [0]{\href }%
\providecommand \doibase [0]{http://dx.doi.org/}%
\providecommand \selectlanguage [0]{\@gobble}%
\providecommand \bibinfo  [0]{\@secondoftwo}%
\providecommand \bibfield  [0]{\@secondoftwo}%
\providecommand \translation [1]{[#1]}%
\providecommand \BibitemOpen [0]{}%
\providecommand \bibitemStop [0]{}%
\providecommand \bibitemNoStop [0]{.\EOS\space}%
\providecommand \EOS [0]{\spacefactor3000\relax}%
\providecommand \BibitemShut  [1]{\csname bibitem#1\endcsname}%
\let\auto@bib@innerbib\@empty
\bibitem [{\citenamefont {Thomas}\ \emph {et~al.}(2014)\citenamefont {Thomas},
  \citenamefont {Jan}, \citenamefont {Zhu}, \citenamefont {Liu}, \citenamefont
  {Lee}, \citenamefont {Le}, \citenamefont {Tong}, \citenamefont {Pi},
  \citenamefont {Wang}, \citenamefont {Shen}, \citenamefont {He}, \citenamefont
  {Haq}, \citenamefont {Teng}, \citenamefont {Lam}, \citenamefont {Huang},
  \citenamefont {Zhong}, \citenamefont {Torng},\ and\ \citenamefont
  {Wang}}]{thomas2014jap}%
  \BibitemOpen
  \bibfield  {author} {\bibinfo {author} {\bibfnamefont {L.}~\bibnamefont
  {Thomas}}, \bibinfo {author} {\bibfnamefont {G.}~\bibnamefont {Jan}},
  \bibinfo {author} {\bibfnamefont {J.}~\bibnamefont {Zhu}}, \bibinfo {author}
  {\bibfnamefont {H.~L.}\ \bibnamefont {Liu}}, \bibinfo {author} {\bibfnamefont
  {Y.~J.}\ \bibnamefont {Lee}}, \bibinfo {author} {\bibfnamefont
  {S.}~\bibnamefont {Le}}, \bibinfo {author} {\bibfnamefont {R.~Y.}\
  \bibnamefont {Tong}}, \bibinfo {author} {\bibfnamefont {K.~Y.}\ \bibnamefont
  {Pi}}, \bibinfo {author} {\bibfnamefont {Y.~J.}\ \bibnamefont {Wang}},
  \bibinfo {author} {\bibfnamefont {D.~N.}\ \bibnamefont {Shen}}, \bibinfo
  {author} {\bibfnamefont {R.~R.}\ \bibnamefont {He}}, \bibinfo {author}
  {\bibfnamefont {J.}~\bibnamefont {Haq}}, \bibinfo {author} {\bibfnamefont
  {J.}~\bibnamefont {Teng}}, \bibinfo {author} {\bibfnamefont {V.}~\bibnamefont
  {Lam}}, \bibinfo {author} {\bibfnamefont {K.~L.}\ \bibnamefont {Huang}},
  \bibinfo {author} {\bibfnamefont {T.}~\bibnamefont {Zhong}}, \bibinfo
  {author} {\bibfnamefont {T.}~\bibnamefont {Torng}}, \ and\ \bibinfo {author}
  {\bibfnamefont {P.~K.}\ \bibnamefont {Wang}},\ }\href@noop {} {\bibfield
  {journal} {\bibinfo  {journal} {J. Appl. Phys.}\ }\textbf {\bibinfo {volume}
  {115}},\ \bibinfo {pages} {172615} (\bibinfo {year} {2014})}\BibitemShut
  {NoStop}%
\bibitem [{\citenamefont {Cubukcu}\ \emph {et~al.}(2014)\citenamefont
  {Cubukcu}, \citenamefont {Boulle}, \citenamefont {Drouard}, \citenamefont
  {Garello}, \citenamefont {Avci}, \citenamefont {Miron}, \citenamefont
  {Langer}, \citenamefont {Ocker}, \citenamefont {Gambardella},\ and\
  \citenamefont {Gaudin}}]{cubukcu2014apl}%
  \BibitemOpen
  \bibfield  {author} {\bibinfo {author} {\bibfnamefont {M.}~\bibnamefont
  {Cubukcu}}, \bibinfo {author} {\bibfnamefont {O.}~\bibnamefont {Boulle}},
  \bibinfo {author} {\bibfnamefont {M.}~\bibnamefont {Drouard}}, \bibinfo
  {author} {\bibfnamefont {K.}~\bibnamefont {Garello}}, \bibinfo {author}
  {\bibfnamefont {C.~O.}\ \bibnamefont {Avci}}, \bibinfo {author}
  {\bibfnamefont {I.~M.}\ \bibnamefont {Miron}}, \bibinfo {author}
  {\bibfnamefont {J.}~\bibnamefont {Langer}}, \bibinfo {author} {\bibfnamefont
  {B.}~\bibnamefont {Ocker}}, \bibinfo {author} {\bibfnamefont
  {P.}~\bibnamefont {Gambardella}}, \ and\ \bibinfo {author} {\bibfnamefont
  {G.}~\bibnamefont {Gaudin}},\ }\href@noop {} {\bibfield  {journal} {\bibinfo
  {journal} {Appl. Phys. Lett.}\ }\textbf {\bibinfo {volume} {104}},\ \bibinfo
  {pages} {042406} (\bibinfo {year} {2014})}\BibitemShut {NoStop}%
\bibitem [{\citenamefont {Fukami}\ \emph {et~al.}(2016)\citenamefont {Fukami},
  \citenamefont {Anekawa}, \citenamefont {Zhang},\ and\ \citenamefont
  {Ohno}}]{fukami2016nnano}%
  \BibitemOpen
  \bibfield  {author} {\bibinfo {author} {\bibfnamefont {S.}~\bibnamefont
  {Fukami}}, \bibinfo {author} {\bibfnamefont {T.}~\bibnamefont {Anekawa}},
  \bibinfo {author} {\bibfnamefont {C.}~\bibnamefont {Zhang}}, \ and\ \bibinfo
  {author} {\bibfnamefont {H.}~\bibnamefont {Ohno}},\ }\href@noop {} {\bibfield
   {journal} {\bibinfo  {journal} {Nat Nanotechnol.}\ }\textbf {\bibinfo
  {volume} {11}},\ \bibinfo {pages} {621} (\bibinfo {year} {2016})}\BibitemShut
  {NoStop}%
\bibitem [{\citenamefont {Parkin}\ \emph {et~al.}(2008)\citenamefont {Parkin},
  \citenamefont {Hayashi},\ and\ \citenamefont {Thomas}}]{parkin2008science}%
  \BibitemOpen
  \bibfield  {author} {\bibinfo {author} {\bibfnamefont {S.~S.~P.}\
  \bibnamefont {Parkin}}, \bibinfo {author} {\bibfnamefont {M.}~\bibnamefont
  {Hayashi}}, \ and\ \bibinfo {author} {\bibfnamefont {L.}~\bibnamefont
  {Thomas}},\ }\href@noop {} {\bibfield  {journal} {\bibinfo  {journal}
  {Science}\ }\textbf {\bibinfo {volume} {320}},\ \bibinfo {pages} {190}
  (\bibinfo {year} {2008})}\BibitemShut {NoStop}%
\bibitem [{\citenamefont {Parkin}\ and\ \citenamefont
  {Yang}(2015)}]{parkin2015nnano}%
  \BibitemOpen
  \bibfield  {author} {\bibinfo {author} {\bibfnamefont {S.}~\bibnamefont
  {Parkin}}\ and\ \bibinfo {author} {\bibfnamefont {S.-H.}\ \bibnamefont
  {Yang}},\ }\href@noop {} {\bibfield  {journal} {\bibinfo  {journal} {Nat.
  Nanotechnol.}\ }\textbf {\bibinfo {volume} {10}},\ \bibinfo {pages} {195}
  (\bibinfo {year} {2015})}\BibitemShut {NoStop}%
\bibitem [{\citenamefont {Mangin}\ \emph {et~al.}(2006)\citenamefont {Mangin},
  \citenamefont {Ravelosona}, \citenamefont {Katine}, \citenamefont {Carey},
  \citenamefont {Terris},\ and\ \citenamefont {Fullerton}}]{mangin2006nmat}%
  \BibitemOpen
  \bibfield  {author} {\bibinfo {author} {\bibfnamefont {S.}~\bibnamefont
  {Mangin}}, \bibinfo {author} {\bibfnamefont {D.}~\bibnamefont {Ravelosona}},
  \bibinfo {author} {\bibfnamefont {J.~A.}\ \bibnamefont {Katine}}, \bibinfo
  {author} {\bibfnamefont {M.~J.}\ \bibnamefont {Carey}}, \bibinfo {author}
  {\bibfnamefont {B.~D.}\ \bibnamefont {Terris}}, \ and\ \bibinfo {author}
  {\bibfnamefont {E.~E.}\ \bibnamefont {Fullerton}},\ }\href@noop {} {\bibfield
   {journal} {\bibinfo  {journal} {Nat. Mater.}\ }\textbf {\bibinfo {volume}
  {5}},\ \bibinfo {pages} {210} (\bibinfo {year} {2006})}\BibitemShut {NoStop}%
\bibitem [{\citenamefont {Ikeda}\ \emph {et~al.}(2010)\citenamefont {Ikeda},
  \citenamefont {Miura}, \citenamefont {Yamamoto}, \citenamefont {Mizunuma},
  \citenamefont {Gan}, \citenamefont {Endo}, \citenamefont {Kanai},
  \citenamefont {Hayakawa}, \citenamefont {Matsukura},\ and\ \citenamefont
  {Ohno}}]{ikeda2010nmat}%
  \BibitemOpen
  \bibfield  {author} {\bibinfo {author} {\bibfnamefont {S.}~\bibnamefont
  {Ikeda}}, \bibinfo {author} {\bibfnamefont {K.}~\bibnamefont {Miura}},
  \bibinfo {author} {\bibfnamefont {H.}~\bibnamefont {Yamamoto}}, \bibinfo
  {author} {\bibfnamefont {K.}~\bibnamefont {Mizunuma}}, \bibinfo {author}
  {\bibfnamefont {H.~D.}\ \bibnamefont {Gan}}, \bibinfo {author} {\bibfnamefont
  {M.}~\bibnamefont {Endo}}, \bibinfo {author} {\bibfnamefont {S.}~\bibnamefont
  {Kanai}}, \bibinfo {author} {\bibfnamefont {J.}~\bibnamefont {Hayakawa}},
  \bibinfo {author} {\bibfnamefont {F.}~\bibnamefont {Matsukura}}, \ and\
  \bibinfo {author} {\bibfnamefont {H.}~\bibnamefont {Ohno}},\ }\href@noop {}
  {\bibfield  {journal} {\bibinfo  {journal} {Nat. Mater.}\ }\textbf {\bibinfo
  {volume} {9}},\ \bibinfo {pages} {721} (\bibinfo {year} {2010})}\BibitemShut
  {NoStop}%
\bibitem [{\citenamefont {Carcia}\ \emph {et~al.}(1985)\citenamefont {Carcia},
  \citenamefont {Meinhaldt},\ and\ \citenamefont {Suna}}]{carcia1985apl}%
  \BibitemOpen
  \bibfield  {author} {\bibinfo {author} {\bibfnamefont {P.~F.}\ \bibnamefont
  {Carcia}}, \bibinfo {author} {\bibfnamefont {A.~D.}\ \bibnamefont
  {Meinhaldt}}, \ and\ \bibinfo {author} {\bibfnamefont {A.}~\bibnamefont
  {Suna}},\ }\href@noop {} {\bibfield  {journal} {\bibinfo  {journal} {Appl.
  Phys. Lett.}\ }\textbf {\bibinfo {volume} {47}},\ \bibinfo {pages} {178}
  (\bibinfo {year} {1985})}\BibitemShut {NoStop}%
\bibitem [{\citenamefont {Denbroeder}\ \emph {et~al.}(1988)\citenamefont
  {Denbroeder}, \citenamefont {Kuiper}, \citenamefont {Vandemosselaer},\ and\
  \citenamefont {Hoving}}]{denbroeder1988prl}%
  \BibitemOpen
  \bibfield  {author} {\bibinfo {author} {\bibfnamefont {F.~J.~A.}\
  \bibnamefont {Denbroeder}}, \bibinfo {author} {\bibfnamefont
  {D.}~\bibnamefont {Kuiper}}, \bibinfo {author} {\bibfnamefont {A.~P.}\
  \bibnamefont {Vandemosselaer}}, \ and\ \bibinfo {author} {\bibfnamefont
  {W.}~\bibnamefont {Hoving}},\ }\href@noop {} {\bibfield  {journal} {\bibinfo
  {journal} {Phys. Rev. Lett.}\ }\textbf {\bibinfo {volume} {60}},\ \bibinfo
  {pages} {2769} (\bibinfo {year} {1988})}\BibitemShut {NoStop}%
\bibitem [{\citenamefont {den Broeder}\ \emph {et~al.}(1991)\citenamefont {den
  Broeder}, \citenamefont {Hoving},\ and\ \citenamefont
  {Bloemen}}]{denBroeder1991jmmm}%
  \BibitemOpen
  \bibfield  {author} {\bibinfo {author} {\bibfnamefont {F.~J.~A.}\
  \bibnamefont {den Broeder}}, \bibinfo {author} {\bibfnamefont
  {W.}~\bibnamefont {Hoving}}, \ and\ \bibinfo {author} {\bibfnamefont
  {P.~J.~H.}\ \bibnamefont {Bloemen}},\ }\href@noop {} {\bibfield  {journal}
  {\bibinfo  {journal} {Journal of Magnetism and Magnetic Materials}\ }\textbf
  {\bibinfo {volume} {93}},\ \bibinfo {pages} {562} (\bibinfo {year}
  {1991})}\BibitemShut {NoStop}%
\bibitem [{\citenamefont {Daalderop}\ \emph {et~al.}(1992)\citenamefont
  {Daalderop}, \citenamefont {Kelly},\ and\ \citenamefont {den
  Broeder}}]{daalderop1992prl}%
  \BibitemOpen
  \bibfield  {author} {\bibinfo {author} {\bibfnamefont {G.~H.~O.}\
  \bibnamefont {Daalderop}}, \bibinfo {author} {\bibfnamefont {P.~J.}\
  \bibnamefont {Kelly}}, \ and\ \bibinfo {author} {\bibfnamefont {F.~J.~A.}\
  \bibnamefont {den Broeder}},\ }\href@noop {} {\bibfield  {journal} {\bibinfo
  {journal} {Phys. Rev. Lett.}\ }\textbf {\bibinfo {volume} {68}},\ \bibinfo
  {pages} {682} (\bibinfo {year} {1992})}\BibitemShut {NoStop}%
\bibitem [{\citenamefont {Johnson}\ \emph {et~al.}(1996)\citenamefont
  {Johnson}, \citenamefont {Bloemen}, \citenamefont {den Broeder},\ and\
  \citenamefont {deVries}}]{johnson1996rpp}%
  \BibitemOpen
  \bibfield  {author} {\bibinfo {author} {\bibfnamefont {M.~T.}\ \bibnamefont
  {Johnson}}, \bibinfo {author} {\bibfnamefont {P.~J.~H.}\ \bibnamefont
  {Bloemen}}, \bibinfo {author} {\bibfnamefont {F.~J.~A.}\ \bibnamefont {den
  Broeder}}, \ and\ \bibinfo {author} {\bibfnamefont {J.~J.}\ \bibnamefont
  {deVries}},\ }\href@noop {} {\bibfield  {journal} {\bibinfo  {journal} {Rep.
  Prog. Phys.}\ }\textbf {\bibinfo {volume} {59}},\ \bibinfo {pages} {1409}
  (\bibinfo {year} {1996})}\BibitemShut {NoStop}%
\bibitem [{\citenamefont {Nakajima}\ \emph {et~al.}(1998)\citenamefont
  {Nakajima}, \citenamefont {Koide}, \citenamefont {Shidara}, \citenamefont
  {Miyauchi}, \citenamefont {Fukutani}, \citenamefont {Fujimori}, \citenamefont
  {Iio}, \citenamefont {Katayama}, \citenamefont {Nyvlt},\ and\ \citenamefont
  {Suzuki}}]{nakajima1998prl}%
  \BibitemOpen
  \bibfield  {author} {\bibinfo {author} {\bibfnamefont {N.}~\bibnamefont
  {Nakajima}}, \bibinfo {author} {\bibfnamefont {T.}~\bibnamefont {Koide}},
  \bibinfo {author} {\bibfnamefont {T.}~\bibnamefont {Shidara}}, \bibinfo
  {author} {\bibfnamefont {H.}~\bibnamefont {Miyauchi}}, \bibinfo {author}
  {\bibfnamefont {I.}~\bibnamefont {Fukutani}}, \bibinfo {author}
  {\bibfnamefont {A.}~\bibnamefont {Fujimori}}, \bibinfo {author}
  {\bibfnamefont {K.}~\bibnamefont {Iio}}, \bibinfo {author} {\bibfnamefont
  {T.}~\bibnamefont {Katayama}}, \bibinfo {author} {\bibfnamefont
  {M.}~\bibnamefont {Nyvlt}}, \ and\ \bibinfo {author} {\bibfnamefont
  {Y.}~\bibnamefont {Suzuki}},\ }\href@noop {} {\bibfield  {journal} {\bibinfo
  {journal} {Phys. Rev. Lett.}\ }\textbf {\bibinfo {volume} {81}},\ \bibinfo
  {pages} {5229} (\bibinfo {year} {1998})}\BibitemShut {NoStop}%
\bibitem [{\citenamefont {Manchon}\ \emph {et~al.}(2008)\citenamefont
  {Manchon}, \citenamefont {Ducruet}, \citenamefont {Lombard}, \citenamefont
  {Auffret}, \citenamefont {Rodmacq}, \citenamefont {Dieny}, \citenamefont
  {Pizzini}, \citenamefont {Vogel}, \citenamefont {Uhlir}, \citenamefont
  {Hochstrasser},\ and\ \citenamefont {Panaccione}}]{manchon2008jap}%
  \BibitemOpen
  \bibfield  {author} {\bibinfo {author} {\bibfnamefont {A.}~\bibnamefont
  {Manchon}}, \bibinfo {author} {\bibfnamefont {C.}~\bibnamefont {Ducruet}},
  \bibinfo {author} {\bibfnamefont {L.}~\bibnamefont {Lombard}}, \bibinfo
  {author} {\bibfnamefont {S.}~\bibnamefont {Auffret}}, \bibinfo {author}
  {\bibfnamefont {B.}~\bibnamefont {Rodmacq}}, \bibinfo {author} {\bibfnamefont
  {B.}~\bibnamefont {Dieny}}, \bibinfo {author} {\bibfnamefont
  {S.}~\bibnamefont {Pizzini}}, \bibinfo {author} {\bibfnamefont
  {J.}~\bibnamefont {Vogel}}, \bibinfo {author} {\bibfnamefont
  {V.}~\bibnamefont {Uhlir}}, \bibinfo {author} {\bibfnamefont
  {M.}~\bibnamefont {Hochstrasser}}, \ and\ \bibinfo {author} {\bibfnamefont
  {G.}~\bibnamefont {Panaccione}},\ }\href@noop {} {\bibfield  {journal}
  {\bibinfo  {journal} {J. Appl. Phys.}\ }\textbf {\bibinfo {volume} {104}},\
  \bibinfo {pages} {043914} (\bibinfo {year} {2008})}\BibitemShut {NoStop}%
\bibitem [{\citenamefont {Yakata}\ \emph {et~al.}(2009)\citenamefont {Yakata},
  \citenamefont {Kubota}, \citenamefont {Suzuki}, \citenamefont {Yakushiji},
  \citenamefont {Fukushima}, \citenamefont {Yuasa},\ and\ \citenamefont
  {Ando}}]{yakata2009jap}%
  \BibitemOpen
  \bibfield  {author} {\bibinfo {author} {\bibfnamefont {S.}~\bibnamefont
  {Yakata}}, \bibinfo {author} {\bibfnamefont {H.}~\bibnamefont {Kubota}},
  \bibinfo {author} {\bibfnamefont {Y.}~\bibnamefont {Suzuki}}, \bibinfo
  {author} {\bibfnamefont {K.}~\bibnamefont {Yakushiji}}, \bibinfo {author}
  {\bibfnamefont {A.}~\bibnamefont {Fukushima}}, \bibinfo {author}
  {\bibfnamefont {S.}~\bibnamefont {Yuasa}}, \ and\ \bibinfo {author}
  {\bibfnamefont {K.}~\bibnamefont {Ando}},\ }\href@noop {} {\bibfield
  {journal} {\bibinfo  {journal} {Journal of Applied Physics}\ }\textbf
  {\bibinfo {volume} {105}},\ \bibinfo {pages} {07D131} (\bibinfo {year}
  {2009})}\BibitemShut {NoStop}%
\bibitem [{\citenamefont {Kubota}\ \emph {et~al.}(2012)\citenamefont {Kubota},
  \citenamefont {Ishibashi}, \citenamefont {Saruya}, \citenamefont {Nozaki},
  \citenamefont {Fukushima}, \citenamefont {Yakushiji}, \citenamefont {Ando},
  \citenamefont {Suzuki},\ and\ \citenamefont {Yuasa}}]{kubota2012jap}%
  \BibitemOpen
  \bibfield  {author} {\bibinfo {author} {\bibfnamefont {H.}~\bibnamefont
  {Kubota}}, \bibinfo {author} {\bibfnamefont {S.}~\bibnamefont {Ishibashi}},
  \bibinfo {author} {\bibfnamefont {T.}~\bibnamefont {Saruya}}, \bibinfo
  {author} {\bibfnamefont {T.}~\bibnamefont {Nozaki}}, \bibinfo {author}
  {\bibfnamefont {A.}~\bibnamefont {Fukushima}}, \bibinfo {author}
  {\bibfnamefont {K.}~\bibnamefont {Yakushiji}}, \bibinfo {author}
  {\bibfnamefont {K.}~\bibnamefont {Ando}}, \bibinfo {author} {\bibfnamefont
  {Y.}~\bibnamefont {Suzuki}}, \ and\ \bibinfo {author} {\bibfnamefont
  {S.}~\bibnamefont {Yuasa}},\ }\href@noop {} {\bibfield  {journal} {\bibinfo
  {journal} {Journal of Applied Physics}\ }\textbf {\bibinfo {volume} {111}},\
  \bibinfo {pages} {07C723} (\bibinfo {year} {2012})}\BibitemShut {NoStop}%
\bibitem [{\citenamefont {Sato}\ \emph {et~al.}(2014)\citenamefont {Sato},
  \citenamefont {Enobio}, \citenamefont {Yamanouchi}, \citenamefont {Ikeda},
  \citenamefont {Fukami}, \citenamefont {Kanai}, \citenamefont {Matsukura},\
  and\ \citenamefont {Ohno}}]{sato2014apl}%
  \BibitemOpen
  \bibfield  {author} {\bibinfo {author} {\bibfnamefont {H.}~\bibnamefont
  {Sato}}, \bibinfo {author} {\bibfnamefont {E.~C.~I.}\ \bibnamefont {Enobio}},
  \bibinfo {author} {\bibfnamefont {M.}~\bibnamefont {Yamanouchi}}, \bibinfo
  {author} {\bibfnamefont {S.}~\bibnamefont {Ikeda}}, \bibinfo {author}
  {\bibfnamefont {S.}~\bibnamefont {Fukami}}, \bibinfo {author} {\bibfnamefont
  {S.}~\bibnamefont {Kanai}}, \bibinfo {author} {\bibfnamefont
  {F.}~\bibnamefont {Matsukura}}, \ and\ \bibinfo {author} {\bibfnamefont
  {H.}~\bibnamefont {Ohno}},\ }\href@noop {} {\bibfield  {journal} {\bibinfo
  {journal} {Appl. Phys. Lett.}\ }\textbf {\bibinfo {volume} {105}},\ \bibinfo
  {pages} {062403} (\bibinfo {year} {2014})}\BibitemShut {NoStop}%
\bibitem [{\citenamefont {Parkin}\ \emph {et~al.}(1990)\citenamefont {Parkin},
  \citenamefont {More},\ and\ \citenamefont {Roche}}]{parkin1990prl}%
  \BibitemOpen
  \bibfield  {author} {\bibinfo {author} {\bibfnamefont {S.~S.~P.}\
  \bibnamefont {Parkin}}, \bibinfo {author} {\bibfnamefont {N.}~\bibnamefont
  {More}}, \ and\ \bibinfo {author} {\bibfnamefont {K.~P.}\ \bibnamefont
  {Roche}},\ }\href@noop {} {\bibfield  {journal} {\bibinfo  {journal} {Phys.
  Rev. Lett.}\ }\textbf {\bibinfo {volume} {64}},\ \bibinfo {pages} {2304}
  (\bibinfo {year} {1990})}\BibitemShut {NoStop}%
\bibitem [{\citenamefont {Parkin}(1991)}]{parkin1991prl}%
  \BibitemOpen
  \bibfield  {author} {\bibinfo {author} {\bibfnamefont {S.~S.~P.}\
  \bibnamefont {Parkin}},\ }\href@noop {} {\bibfield  {journal} {\bibinfo
  {journal} {Phys. Rev. Lett.}\ }\textbf {\bibinfo {volume} {67}},\ \bibinfo
  {pages} {3598} (\bibinfo {year} {1991})}\BibitemShut {NoStop}%
\bibitem [{\citenamefont {Yang}\ \emph {et~al.}(2015)\citenamefont {Yang},
  \citenamefont {Ryu},\ and\ \citenamefont {Parkin}}]{yang2015nnano}%
  \BibitemOpen
  \bibfield  {author} {\bibinfo {author} {\bibfnamefont {S.-H.}\ \bibnamefont
  {Yang}}, \bibinfo {author} {\bibfnamefont {K.-S.}\ \bibnamefont {Ryu}}, \
  and\ \bibinfo {author} {\bibfnamefont {S.}~\bibnamefont {Parkin}},\
  }\href@noop {} {\bibfield  {journal} {\bibinfo  {journal} {Nat.
  Nanotechnol.}\ }\textbf {\bibinfo {volume} {10}},\ \bibinfo {pages} {221}
  (\bibinfo {year} {2015})}\BibitemShut {NoStop}%
\bibitem [{\citenamefont {Lau}\ \emph {et~al.}(2016)\citenamefont {Lau},
  \citenamefont {Betto}, \citenamefont {Rode}, \citenamefont {Coey},\ and\
  \citenamefont {Stamenov}}]{lau2016nnano}%
  \BibitemOpen
  \bibfield  {author} {\bibinfo {author} {\bibfnamefont {Y.-C.}\ \bibnamefont
  {Lau}}, \bibinfo {author} {\bibfnamefont {D.}~\bibnamefont {Betto}}, \bibinfo
  {author} {\bibfnamefont {K.}~\bibnamefont {Rode}}, \bibinfo {author}
  {\bibfnamefont {J.~M.~D.}\ \bibnamefont {Coey}}, \ and\ \bibinfo {author}
  {\bibfnamefont {P.}~\bibnamefont {Stamenov}},\ }\href@noop {} {\bibfield
  {journal} {\bibinfo  {journal} {Nat Nano}\ }\textbf {\bibinfo {volume}
  {11}},\ \bibinfo {pages} {758} (\bibinfo {year} {2016})}\BibitemShut
  {NoStop}%
\bibitem [{\citenamefont {Roschewsky}\ \emph {et~al.}(2016)\citenamefont
  {Roschewsky}, \citenamefont {Matsumura}, \citenamefont {Cheema},
  \citenamefont {Hellman}, \citenamefont {Kato}, \citenamefont {Iwata},\ and\
  \citenamefont {Salahuddin}}]{roschewsky2016apl}%
  \BibitemOpen
  \bibfield  {author} {\bibinfo {author} {\bibfnamefont {N.}~\bibnamefont
  {Roschewsky}}, \bibinfo {author} {\bibfnamefont {T.}~\bibnamefont
  {Matsumura}}, \bibinfo {author} {\bibfnamefont {S.}~\bibnamefont {Cheema}},
  \bibinfo {author} {\bibfnamefont {F.}~\bibnamefont {Hellman}}, \bibinfo
  {author} {\bibfnamefont {T.}~\bibnamefont {Kato}}, \bibinfo {author}
  {\bibfnamefont {S.}~\bibnamefont {Iwata}}, \ and\ \bibinfo {author}
  {\bibfnamefont {S.}~\bibnamefont {Salahuddin}},\ }\href@noop {} {\bibfield
  {journal} {\bibinfo  {journal} {Applied Physics Letters}\ }\textbf {\bibinfo
  {volume} {109}},\ \bibinfo {pages} {112403} (\bibinfo {year}
  {2016})}\BibitemShut {NoStop}%
\bibitem [{\citenamefont {Finley}\ and\ \citenamefont
  {Liu}(2016)}]{finley2016prap}%
  \BibitemOpen
  \bibfield  {author} {\bibinfo {author} {\bibfnamefont {J.}~\bibnamefont
  {Finley}}\ and\ \bibinfo {author} {\bibfnamefont {L.}~\bibnamefont {Liu}},\
  }\href@noop {} {\bibfield  {journal} {\bibinfo  {journal} {Physical Review
  Applied}\ }\textbf {\bibinfo {volume} {6}},\ \bibinfo {pages} {054001}
  (\bibinfo {year} {2016})}\BibitemShut {NoStop}%
\bibitem [{\citenamefont {Mishra}\ \emph {et~al.}(2017)\citenamefont {Mishra},
  \citenamefont {Yu}, \citenamefont {Qiu}, \citenamefont {Motapothula},
  \citenamefont {Venkatesan},\ and\ \citenamefont {Yang}}]{mishra2017prl}%
  \BibitemOpen
  \bibfield  {author} {\bibinfo {author} {\bibfnamefont {R.}~\bibnamefont
  {Mishra}}, \bibinfo {author} {\bibfnamefont {J.}~\bibnamefont {Yu}}, \bibinfo
  {author} {\bibfnamefont {X.}~\bibnamefont {Qiu}}, \bibinfo {author}
  {\bibfnamefont {M.}~\bibnamefont {Motapothula}}, \bibinfo {author}
  {\bibfnamefont {T.}~\bibnamefont {Venkatesan}}, \ and\ \bibinfo {author}
  {\bibfnamefont {H.}~\bibnamefont {Yang}},\ }\href@noop {} {\bibfield
  {journal} {\bibinfo  {journal} {Physical Review Letters}\ }\textbf {\bibinfo
  {volume} {118}},\ \bibinfo {pages} {167201} (\bibinfo {year}
  {2017})}\BibitemShut {NoStop}%
\bibitem [{\citenamefont {Blasing}\ \emph {et~al.}(2018)\citenamefont
  {Blasing}, \citenamefont {Ma}, \citenamefont {Yang}, \citenamefont {Garg},
  \citenamefont {Dejene}, \citenamefont {N'Diaye}, \citenamefont {Chen},
  \citenamefont {Liu},\ and\ \citenamefont {Parkin}}]{blasing2018ncomm}%
  \BibitemOpen
  \bibfield  {author} {\bibinfo {author} {\bibfnamefont {R.}~\bibnamefont
  {Blasing}}, \bibinfo {author} {\bibfnamefont {T.~P.}\ \bibnamefont {Ma}},
  \bibinfo {author} {\bibfnamefont {S.~H.}\ \bibnamefont {Yang}}, \bibinfo
  {author} {\bibfnamefont {C.}~\bibnamefont {Garg}}, \bibinfo {author}
  {\bibfnamefont {F.~K.}\ \bibnamefont {Dejene}}, \bibinfo {author}
  {\bibfnamefont {A.~T.}\ \bibnamefont {N'Diaye}}, \bibinfo {author}
  {\bibfnamefont {G.}~\bibnamefont {Chen}}, \bibinfo {author} {\bibfnamefont
  {K.}~\bibnamefont {Liu}}, \ and\ \bibinfo {author} {\bibfnamefont {S.~S.~P.}\
  \bibnamefont {Parkin}},\ }\href@noop {} {\bibfield  {journal} {\bibinfo
  {journal} {Nature Communications}\ }\textbf {\bibinfo {volume} {9}},\
  \bibinfo {pages} {4984} (\bibinfo {year} {2018})}\BibitemShut {NoStop}%
\bibitem [{\citenamefont {Bruno}(1989)}]{bruno1989prb}%
  \BibitemOpen
  \bibfield  {author} {\bibinfo {author} {\bibfnamefont {P.}~\bibnamefont
  {Bruno}},\ }\href@noop {} {\bibfield  {journal} {\bibinfo  {journal}
  {Physical Review B}\ }\textbf {\bibinfo {volume} {39}},\ \bibinfo {pages}
  {865} (\bibinfo {year} {1989})}\BibitemShut {NoStop}%
\bibitem [{\citenamefont {Nakamura}\ \emph {et~al.}(2009)\citenamefont
  {Nakamura}, \citenamefont {Shimabukuro}, \citenamefont {Fujiwara},
  \citenamefont {Akiyama}, \citenamefont {Ito},\ and\ \citenamefont
  {Freeman}}]{nakamura2009prl}%
  \BibitemOpen
  \bibfield  {author} {\bibinfo {author} {\bibfnamefont {K.}~\bibnamefont
  {Nakamura}}, \bibinfo {author} {\bibfnamefont {R.}~\bibnamefont
  {Shimabukuro}}, \bibinfo {author} {\bibfnamefont {Y.}~\bibnamefont
  {Fujiwara}}, \bibinfo {author} {\bibfnamefont {T.}~\bibnamefont {Akiyama}},
  \bibinfo {author} {\bibfnamefont {T.}~\bibnamefont {Ito}}, \ and\ \bibinfo
  {author} {\bibfnamefont {A.~J.}\ \bibnamefont {Freeman}},\ }\href@noop {}
  {\bibfield  {journal} {\bibinfo  {journal} {Phys. Rev. Lett.}\ }\textbf
  {\bibinfo {volume} {102}},\ \bibinfo {pages} {187201} (\bibinfo {year}
  {2009})}\BibitemShut {NoStop}%
\bibitem [{\citenamefont {Yang}\ \emph {et~al.}(2011)\citenamefont {Yang},
  \citenamefont {Chshiev}, \citenamefont {Dieny}, \citenamefont {Lee},
  \citenamefont {Manchon},\ and\ \citenamefont {Shin}}]{yang2011prb}%
  \BibitemOpen
  \bibfield  {author} {\bibinfo {author} {\bibfnamefont {H.~X.}\ \bibnamefont
  {Yang}}, \bibinfo {author} {\bibfnamefont {M.}~\bibnamefont {Chshiev}},
  \bibinfo {author} {\bibfnamefont {B.}~\bibnamefont {Dieny}}, \bibinfo
  {author} {\bibfnamefont {J.~H.}\ \bibnamefont {Lee}}, \bibinfo {author}
  {\bibfnamefont {A.}~\bibnamefont {Manchon}}, \ and\ \bibinfo {author}
  {\bibfnamefont {K.~H.}\ \bibnamefont {Shin}},\ }\href@noop {} {\bibfield
  {journal} {\bibinfo  {journal} {Phys. Rev. B}\ }\textbf {\bibinfo {volume}
  {84}},\ \bibinfo {pages} {054401} (\bibinfo {year} {2011})}\BibitemShut
  {NoStop}%
\bibitem [{\citenamefont {Dieny}\ and\ \citenamefont
  {Chshiev}(2017)}]{dieny2017rmp}%
  \BibitemOpen
  \bibfield  {author} {\bibinfo {author} {\bibfnamefont {B.}~\bibnamefont
  {Dieny}}\ and\ \bibinfo {author} {\bibfnamefont {M.}~\bibnamefont
  {Chshiev}},\ }\href@noop {} {\bibfield  {journal} {\bibinfo  {journal}
  {Reviews of Modern Physics}\ }\textbf {\bibinfo {volume} {89}},\ \bibinfo
  {pages} {025008} (\bibinfo {year} {2017})}\BibitemShut {NoStop}%
\bibitem [{\citenamefont {Barnes}\ \emph {et~al.}(2014)\citenamefont {Barnes},
  \citenamefont {Ieda},\ and\ \citenamefont {Maekawa}}]{barnes2014scirep}%
  \BibitemOpen
  \bibfield  {author} {\bibinfo {author} {\bibfnamefont {S.~E.}\ \bibnamefont
  {Barnes}}, \bibinfo {author} {\bibfnamefont {J.}~\bibnamefont {Ieda}}, \ and\
  \bibinfo {author} {\bibfnamefont {S.}~\bibnamefont {Maekawa}},\ }\href@noop
  {} {\bibfield  {journal} {\bibinfo  {journal} {Scientific Reports}\ }\textbf
  {\bibinfo {volume} {4}},\ \bibinfo {pages} {4105} (\bibinfo {year}
  {2014})}\BibitemShut {NoStop}%
\bibitem [{\citenamefont {Kim}\ \emph {et~al.}(2016{\natexlab{a}})\citenamefont
  {Kim}, \citenamefont {Lee}, \citenamefont {Lee},\ and\ \citenamefont
  {Stiles}}]{kim2016prb}%
  \BibitemOpen
  \bibfield  {author} {\bibinfo {author} {\bibfnamefont {K.~W.}\ \bibnamefont
  {Kim}}, \bibinfo {author} {\bibfnamefont {K.~J.}\ \bibnamefont {Lee}},
  \bibinfo {author} {\bibfnamefont {H.~W.}\ \bibnamefont {Lee}}, \ and\
  \bibinfo {author} {\bibfnamefont {M.~D.}\ \bibnamefont {Stiles}},\
  }\href@noop {} {\bibfield  {journal} {\bibinfo  {journal} {Phys. Rev. B}\
  }\textbf {\bibinfo {volume} {94}},\ \bibinfo {pages} {184402} (\bibinfo
  {year} {2016}{\natexlab{a}})}\BibitemShut {NoStop}%
\bibitem [{\citenamefont {Kim}\ \emph {et~al.}(2013)\citenamefont {Kim},
  \citenamefont {Lee}, \citenamefont {Lee},\ and\ \citenamefont
  {Stiles}}]{kim2013prl}%
  \BibitemOpen
  \bibfield  {author} {\bibinfo {author} {\bibfnamefont {K.~W.}\ \bibnamefont
  {Kim}}, \bibinfo {author} {\bibfnamefont {H.~W.}\ \bibnamefont {Lee}},
  \bibinfo {author} {\bibfnamefont {K.~J.}\ \bibnamefont {Lee}}, \ and\
  \bibinfo {author} {\bibfnamefont {M.~D.}\ \bibnamefont {Stiles}},\
  }\href@noop {} {\bibfield  {journal} {\bibinfo  {journal} {Phys. Rev. Lett.}\
  }\textbf {\bibinfo {volume} {111}},\ \bibinfo {pages} {216601} (\bibinfo
  {year} {2013})}\BibitemShut {NoStop}%
\bibitem [{\citenamefont {Manchon}\ \emph {et~al.}(2015)\citenamefont
  {Manchon}, \citenamefont {Koo}, \citenamefont {Nitta}, \citenamefont
  {Frolov},\ and\ \citenamefont {Duine}}]{manchon2015nmat}%
  \BibitemOpen
  \bibfield  {author} {\bibinfo {author} {\bibfnamefont {A.}~\bibnamefont
  {Manchon}}, \bibinfo {author} {\bibfnamefont {H.~C.}\ \bibnamefont {Koo}},
  \bibinfo {author} {\bibfnamefont {J.}~\bibnamefont {Nitta}}, \bibinfo
  {author} {\bibfnamefont {S.~M.}\ \bibnamefont {Frolov}}, \ and\ \bibinfo
  {author} {\bibfnamefont {R.~A.}\ \bibnamefont {Duine}},\ }\href@noop {}
  {\bibfield  {journal} {\bibinfo  {journal} {Nat. Mater.}\ }\textbf {\bibinfo
  {volume} {14}},\ \bibinfo {pages} {871} (\bibinfo {year} {2015})}\BibitemShut
  {NoStop}%
\bibitem [{\citenamefont {Nakayama}\ \emph {et~al.}(2013)\citenamefont
  {Nakayama}, \citenamefont {Althammer}, \citenamefont {Chen}, \citenamefont
  {Uchida}, \citenamefont {Kajiwara}, \citenamefont {Kikuchi}, \citenamefont
  {Ohtani}, \citenamefont {Geprags}, \citenamefont {Opel}, \citenamefont
  {Takahashi}, \citenamefont {Gross}, \citenamefont {Bauer}, \citenamefont
  {Goennenwein},\ and\ \citenamefont {Saitoh}}]{nakayama2013prl}%
  \BibitemOpen
  \bibfield  {author} {\bibinfo {author} {\bibfnamefont {H.}~\bibnamefont
  {Nakayama}}, \bibinfo {author} {\bibfnamefont {M.}~\bibnamefont {Althammer}},
  \bibinfo {author} {\bibfnamefont {Y.~T.}\ \bibnamefont {Chen}}, \bibinfo
  {author} {\bibfnamefont {K.}~\bibnamefont {Uchida}}, \bibinfo {author}
  {\bibfnamefont {Y.}~\bibnamefont {Kajiwara}}, \bibinfo {author}
  {\bibfnamefont {D.}~\bibnamefont {Kikuchi}}, \bibinfo {author} {\bibfnamefont
  {T.}~\bibnamefont {Ohtani}}, \bibinfo {author} {\bibfnamefont
  {S.}~\bibnamefont {Geprags}}, \bibinfo {author} {\bibfnamefont
  {M.}~\bibnamefont {Opel}}, \bibinfo {author} {\bibfnamefont {S.}~\bibnamefont
  {Takahashi}}, \bibinfo {author} {\bibfnamefont {R.}~\bibnamefont {Gross}},
  \bibinfo {author} {\bibfnamefont {G.~E.~W.}\ \bibnamefont {Bauer}}, \bibinfo
  {author} {\bibfnamefont {S.~T.~B.}\ \bibnamefont {Goennenwein}}, \ and\
  \bibinfo {author} {\bibfnamefont {E.}~\bibnamefont {Saitoh}},\ }\href@noop {}
  {\bibfield  {journal} {\bibinfo  {journal} {Phys. Rev. Lett.}\ }\textbf
  {\bibinfo {volume} {110}},\ \bibinfo {pages} {206601} (\bibinfo {year}
  {2013})}\BibitemShut {NoStop}%
\bibitem [{\citenamefont {Chen}\ \emph {et~al.}(2013)\citenamefont {Chen},
  \citenamefont {Takahashi}, \citenamefont {Nakayama}, \citenamefont
  {Althammer}, \citenamefont {Goennenwein}, \citenamefont {Saitoh},\ and\
  \citenamefont {Bauer}}]{chen2013prb}%
  \BibitemOpen
  \bibfield  {author} {\bibinfo {author} {\bibfnamefont {Y.~T.}\ \bibnamefont
  {Chen}}, \bibinfo {author} {\bibfnamefont {S.}~\bibnamefont {Takahashi}},
  \bibinfo {author} {\bibfnamefont {H.}~\bibnamefont {Nakayama}}, \bibinfo
  {author} {\bibfnamefont {M.}~\bibnamefont {Althammer}}, \bibinfo {author}
  {\bibfnamefont {S.~T.~B.}\ \bibnamefont {Goennenwein}}, \bibinfo {author}
  {\bibfnamefont {E.}~\bibnamefont {Saitoh}}, \ and\ \bibinfo {author}
  {\bibfnamefont {G.~E.~W.}\ \bibnamefont {Bauer}},\ }\href@noop {} {\bibfield
  {journal} {\bibinfo  {journal} {Phys. Rev. B}\ }\textbf {\bibinfo {volume}
  {87}},\ \bibinfo {pages} {144411} (\bibinfo {year} {2013})}\BibitemShut
  {NoStop}%
\bibitem [{\citenamefont {Kim}\ \emph {et~al.}(2016{\natexlab{b}})\citenamefont
  {Kim}, \citenamefont {Sheng}, \citenamefont {Takahashi}, \citenamefont
  {Mitani},\ and\ \citenamefont {Hayashi}}]{kim2016prl}%
  \BibitemOpen
  \bibfield  {author} {\bibinfo {author} {\bibfnamefont {J.}~\bibnamefont
  {Kim}}, \bibinfo {author} {\bibfnamefont {P.}~\bibnamefont {Sheng}}, \bibinfo
  {author} {\bibfnamefont {S.}~\bibnamefont {Takahashi}}, \bibinfo {author}
  {\bibfnamefont {S.}~\bibnamefont {Mitani}}, \ and\ \bibinfo {author}
  {\bibfnamefont {M.}~\bibnamefont {Hayashi}},\ }\href@noop {} {\bibfield
  {journal} {\bibinfo  {journal} {Phys. Rev. Lett.}\ }\textbf {\bibinfo
  {volume} {116}},\ \bibinfo {pages} {097201} (\bibinfo {year}
  {2016}{\natexlab{b}})}\BibitemShut {NoStop}%
\bibitem [{\citenamefont {Ishikuro}\ \emph {et~al.}(2019)\citenamefont
  {Ishikuro}, \citenamefont {Kawaguchi}, \citenamefont {Kato}, \citenamefont
  {Lau},\ and\ \citenamefont {Hayashi}}]{ishikuro2019prb}%
  \BibitemOpen
  \bibfield  {author} {\bibinfo {author} {\bibfnamefont {Y.}~\bibnamefont
  {Ishikuro}}, \bibinfo {author} {\bibfnamefont {M.}~\bibnamefont {Kawaguchi}},
  \bibinfo {author} {\bibfnamefont {N.}~\bibnamefont {Kato}}, \bibinfo {author}
  {\bibfnamefont {Y.-C.}\ \bibnamefont {Lau}}, \ and\ \bibinfo {author}
  {\bibfnamefont {M.}~\bibnamefont {Hayashi}},\ }\href {\doibase
  10.1103/PhysRevB.99.134421} {\bibfield  {journal} {\bibinfo  {journal} {Phys.
  Rev. B}\ }\textbf {\bibinfo {volume} {99}},\ \bibinfo {pages} {134421}
  (\bibinfo {year} {2019})}\BibitemShut {NoStop}%
\bibitem [{\citenamefont {Bandiera}\ \emph {et~al.}(2011)\citenamefont
  {Bandiera}, \citenamefont {Sousa}, \citenamefont {Rodmacq},\ and\
  \citenamefont {Dieny}}]{bandiera2011iml}%
  \BibitemOpen
  \bibfield  {author} {\bibinfo {author} {\bibfnamefont {S.}~\bibnamefont
  {Bandiera}}, \bibinfo {author} {\bibfnamefont {R.~C.}\ \bibnamefont {Sousa}},
  \bibinfo {author} {\bibfnamefont {B.}~\bibnamefont {Rodmacq}}, \ and\
  \bibinfo {author} {\bibfnamefont {B.}~\bibnamefont {Dieny}},\ }\href@noop {}
  {\bibfield  {journal} {\bibinfo  {journal} {IEEE Magnetics Letters}\ }\textbf
  {\bibinfo {volume} {2}},\ \bibinfo {pages} {3000504} (\bibinfo {year}
  {2011})}\BibitemShut {NoStop}%
\bibitem [{\citenamefont {Lee}\ \emph {et~al.}(2014)\citenamefont {Lee},
  \citenamefont {Won}, \citenamefont {Son}, \citenamefont {Lim},\ and\
  \citenamefont {Lee}}]{lee2014iml}%
  \BibitemOpen
  \bibfield  {author} {\bibinfo {author} {\bibfnamefont {T.~Y.}\ \bibnamefont
  {Lee}}, \bibinfo {author} {\bibfnamefont {Y.~C.}\ \bibnamefont {Won}},
  \bibinfo {author} {\bibfnamefont {D.~S.}\ \bibnamefont {Son}}, \bibinfo
  {author} {\bibfnamefont {S.~H.}\ \bibnamefont {Lim}}, \ and\ \bibinfo
  {author} {\bibfnamefont {S.~R.}\ \bibnamefont {Lee}},\ }\href@noop {}
  {\bibfield  {journal} {\bibinfo  {journal} {IEEE Magn. Lett.}\ }\textbf
  {\bibinfo {volume} {5}},\ \bibinfo {pages} {1000104} (\bibinfo {year}
  {2014})}\BibitemShut {NoStop}%
\bibitem [{\citenamefont {Yakushiji}\ \emph {et~al.}(2017)\citenamefont
  {Yakushiji}, \citenamefont {Sugihara}, \citenamefont {Fukushima},
  \citenamefont {Kubota},\ and\ \citenamefont {Yuasa}}]{yakushiji2017apl}%
  \BibitemOpen
  \bibfield  {author} {\bibinfo {author} {\bibfnamefont {K.}~\bibnamefont
  {Yakushiji}}, \bibinfo {author} {\bibfnamefont {A.}~\bibnamefont {Sugihara}},
  \bibinfo {author} {\bibfnamefont {A.}~\bibnamefont {Fukushima}}, \bibinfo
  {author} {\bibfnamefont {H.}~\bibnamefont {Kubota}}, \ and\ \bibinfo {author}
  {\bibfnamefont {S.}~\bibnamefont {Yuasa}},\ }\href@noop {} {\bibfield
  {journal} {\bibinfo  {journal} {Applied Physics Letters}\ }\textbf {\bibinfo
  {volume} {110}},\ \bibinfo {pages} {092406} (\bibinfo {year}
  {2017})}\BibitemShut {NoStop}%
\bibitem [{\citenamefont {Itoh}\ \emph {et~al.}(2003)\citenamefont {Itoh},
  \citenamefont {Yanagihara}, \citenamefont {Suzuki},\ and\ \citenamefont
  {Kita}}]{itoh2003jmmm}%
  \BibitemOpen
  \bibfield  {author} {\bibinfo {author} {\bibfnamefont {H.}~\bibnamefont
  {Itoh}}, \bibinfo {author} {\bibfnamefont {H.}~\bibnamefont {Yanagihara}},
  \bibinfo {author} {\bibfnamefont {K.}~\bibnamefont {Suzuki}}, \ and\ \bibinfo
  {author} {\bibfnamefont {E.}~\bibnamefont {Kita}},\ }\href@noop {} {\bibfield
   {journal} {\bibinfo  {journal} {Journal of Magnetism and Magnetic
  Materials}\ }\textbf {\bibinfo {volume} {257}},\ \bibinfo {pages} {184}
  (\bibinfo {year} {2003})}\BibitemShut {NoStop}%
\bibitem [{\citenamefont {Liu}\ \emph {et~al.}(2004)\citenamefont {Liu},
  \citenamefont {Yue}, \citenamefont {Keavney},\ and\ \citenamefont
  {Adenwalla}}]{liu2004prb}%
  \BibitemOpen
  \bibfield  {author} {\bibinfo {author} {\bibfnamefont {Z.~Y.}\ \bibnamefont
  {Liu}}, \bibinfo {author} {\bibfnamefont {L.~P.}\ \bibnamefont {Yue}},
  \bibinfo {author} {\bibfnamefont {D.~J.}\ \bibnamefont {Keavney}}, \ and\
  \bibinfo {author} {\bibfnamefont {S.}~\bibnamefont {Adenwalla}},\ }\href@noop
  {} {\bibfield  {journal} {\bibinfo  {journal} {Physical Review B}\ }\textbf
  {\bibinfo {volume} {70}},\ \bibinfo {pages} {224423} (\bibinfo {year}
  {2004})}\BibitemShut {NoStop}%
\bibitem [{\citenamefont {Knepper}\ and\ \citenamefont
  {Yang}(2005)}]{knepper2005prb}%
  \BibitemOpen
  \bibfield  {author} {\bibinfo {author} {\bibfnamefont {J.~W.}\ \bibnamefont
  {Knepper}}\ and\ \bibinfo {author} {\bibfnamefont {F.~Y.}\ \bibnamefont
  {Yang}},\ }\href@noop {} {\bibfield  {journal} {\bibinfo  {journal} {Physical
  Review B}\ }\textbf {\bibinfo {volume} {71}},\ \bibinfo {pages} {224403}
  (\bibinfo {year} {2005})}\BibitemShut {NoStop}%
\bibitem [{\citenamefont {Zhao}\ \emph {et~al.}(2008)\citenamefont {Zhao},
  \citenamefont {Wang}, \citenamefont {Liu}, \citenamefont {Han},\ and\
  \citenamefont {Zhang}}]{zhao2008jap}%
  \BibitemOpen
  \bibfield  {author} {\bibinfo {author} {\bibfnamefont {J.}~\bibnamefont
  {Zhao}}, \bibinfo {author} {\bibfnamefont {Y.~J.}\ \bibnamefont {Wang}},
  \bibinfo {author} {\bibfnamefont {Y.~Z.}\ \bibnamefont {Liu}}, \bibinfo
  {author} {\bibfnamefont {X.~F.}\ \bibnamefont {Han}}, \ and\ \bibinfo
  {author} {\bibfnamefont {Z.}~\bibnamefont {Zhang}},\ }\href@noop {}
  {\bibfield  {journal} {\bibinfo  {journal} {Journal of Applied Physics}\
  }\textbf {\bibinfo {volume} {104}},\ \bibinfo {pages} {023911} (\bibinfo
  {year} {2008})}\BibitemShut {NoStop}%
\bibitem [{\citenamefont {Bandiera}\ \emph {et~al.}(2012)\citenamefont
  {Bandiera}, \citenamefont {Sousa}, \citenamefont {Auffret}, \citenamefont
  {Rodmacq},\ and\ \citenamefont {Dieny}}]{bandiera2012apl}%
  \BibitemOpen
  \bibfield  {author} {\bibinfo {author} {\bibfnamefont {S.}~\bibnamefont
  {Bandiera}}, \bibinfo {author} {\bibfnamefont {R.~C.}\ \bibnamefont {Sousa}},
  \bibinfo {author} {\bibfnamefont {S.}~\bibnamefont {Auffret}}, \bibinfo
  {author} {\bibfnamefont {B.}~\bibnamefont {Rodmacq}}, \ and\ \bibinfo
  {author} {\bibfnamefont {B.}~\bibnamefont {Dieny}},\ }\href@noop {}
  {\bibfield  {journal} {\bibinfo  {journal} {Appl. Phys. Lett.}\ }\textbf
  {\bibinfo {volume} {101}},\ \bibinfo {pages} {072410} (\bibinfo {year}
  {2012})}\BibitemShut {NoStop}%
\bibitem [{\citenamefont {Engel}\ \emph {et~al.}(2005)\citenamefont {Engel},
  \citenamefont {Akerman}, \citenamefont {Butcher}, \citenamefont {Dave},
  \citenamefont {DeHerrera}, \citenamefont {Durlam}, \citenamefont
  {Grynkewich}, \citenamefont {Janesky}, \citenamefont {Pietambaram},
  \citenamefont {Rizzo}, \citenamefont {Slaughter}, \citenamefont {Smith},
  \citenamefont {Sun},\ and\ \citenamefont {Tehrani}}]{engel2005ieee}%
  \BibitemOpen
  \bibfield  {author} {\bibinfo {author} {\bibfnamefont {B.~N.}\ \bibnamefont
  {Engel}}, \bibinfo {author} {\bibfnamefont {J.}~\bibnamefont {Akerman}},
  \bibinfo {author} {\bibfnamefont {B.}~\bibnamefont {Butcher}}, \bibinfo
  {author} {\bibfnamefont {R.~W.}\ \bibnamefont {Dave}}, \bibinfo {author}
  {\bibfnamefont {M.}~\bibnamefont {DeHerrera}}, \bibinfo {author}
  {\bibfnamefont {M.}~\bibnamefont {Durlam}}, \bibinfo {author} {\bibfnamefont
  {G.}~\bibnamefont {Grynkewich}}, \bibinfo {author} {\bibfnamefont
  {J.}~\bibnamefont {Janesky}}, \bibinfo {author} {\bibfnamefont {S.~V.}\
  \bibnamefont {Pietambaram}}, \bibinfo {author} {\bibfnamefont {N.~D.}\
  \bibnamefont {Rizzo}}, \bibinfo {author} {\bibfnamefont {J.~M.}\ \bibnamefont
  {Slaughter}}, \bibinfo {author} {\bibfnamefont {K.}~\bibnamefont {Smith}},
  \bibinfo {author} {\bibfnamefont {J.~J.}\ \bibnamefont {Sun}}, \ and\
  \bibinfo {author} {\bibfnamefont {S.}~\bibnamefont {Tehrani}},\ }\href@noop
  {} {\bibfield  {journal} {\bibinfo  {journal} {IEEE Trans. Magn.}\ }\textbf
  {\bibinfo {volume} {41}},\ \bibinfo {pages} {132} (\bibinfo {year}
  {2005})}\BibitemShut {NoStop}%
\bibitem [{\citenamefont {Johnson}\ \emph {et~al.}(1995)\citenamefont
  {Johnson}, \citenamefont {Jungblut}, \citenamefont {Kelly},\ and\
  \citenamefont {Denbroeder}}]{johnson1995jmmm}%
  \BibitemOpen
  \bibfield  {author} {\bibinfo {author} {\bibfnamefont {M.~T.}\ \bibnamefont
  {Johnson}}, \bibinfo {author} {\bibfnamefont {R.}~\bibnamefont {Jungblut}},
  \bibinfo {author} {\bibfnamefont {P.~J.}\ \bibnamefont {Kelly}}, \ and\
  \bibinfo {author} {\bibfnamefont {F.~J.~A.}\ \bibnamefont {Denbroeder}},\
  }\href@noop {} {\bibfield  {journal} {\bibinfo  {journal} {J. Magn. Magn.
  Mater.}\ }\textbf {\bibinfo {volume} {148}},\ \bibinfo {pages} {118}
  (\bibinfo {year} {1995})}\BibitemShut {NoStop}%
\bibitem [{\citenamefont {Sinha}\ \emph {et~al.}(2013)\citenamefont {Sinha},
  \citenamefont {Hayashi}, \citenamefont {Kellock}, \citenamefont {Fukami},
  \citenamefont {Yamanouchi}, \citenamefont {Sato}, \citenamefont {Ikeda},
  \citenamefont {Mitani}, \citenamefont {Yang}, \citenamefont {Parkin},\ and\
  \citenamefont {Ohno}}]{sinha2013apl}%
  \BibitemOpen
  \bibfield  {author} {\bibinfo {author} {\bibfnamefont {J.}~\bibnamefont
  {Sinha}}, \bibinfo {author} {\bibfnamefont {M.}~\bibnamefont {Hayashi}},
  \bibinfo {author} {\bibfnamefont {A.~J.}\ \bibnamefont {Kellock}}, \bibinfo
  {author} {\bibfnamefont {S.}~\bibnamefont {Fukami}}, \bibinfo {author}
  {\bibfnamefont {M.}~\bibnamefont {Yamanouchi}}, \bibinfo {author}
  {\bibfnamefont {M.}~\bibnamefont {Sato}}, \bibinfo {author} {\bibfnamefont
  {S.}~\bibnamefont {Ikeda}}, \bibinfo {author} {\bibfnamefont
  {S.}~\bibnamefont {Mitani}}, \bibinfo {author} {\bibfnamefont {S.~H.}\
  \bibnamefont {Yang}}, \bibinfo {author} {\bibfnamefont {S.~S.~P.}\
  \bibnamefont {Parkin}}, \ and\ \bibinfo {author} {\bibfnamefont
  {H.}~\bibnamefont {Ohno}},\ }\href@noop {} {\bibfield  {journal} {\bibinfo
  {journal} {Appl. Phys. Lett.}\ }\textbf {\bibinfo {volume} {102}},\ \bibinfo
  {pages} {242405} (\bibinfo {year} {2013})}\BibitemShut {NoStop}%
\bibitem [{\citenamefont {Gowtham}\ \emph {et~al.}(2016)\citenamefont
  {Gowtham}, \citenamefont {Stiehl}, \citenamefont {Ralph},\ and\ \citenamefont
  {Buhrman}}]{gowtham2016prb}%
  \BibitemOpen
  \bibfield  {author} {\bibinfo {author} {\bibfnamefont {P.~G.}\ \bibnamefont
  {Gowtham}}, \bibinfo {author} {\bibfnamefont {G.~M.}\ \bibnamefont {Stiehl}},
  \bibinfo {author} {\bibfnamefont {D.~C.}\ \bibnamefont {Ralph}}, \ and\
  \bibinfo {author} {\bibfnamefont {R.~A.}\ \bibnamefont {Buhrman}},\
  }\href@noop {} {\bibfield  {journal} {\bibinfo  {journal} {Physical Review
  B}\ }\textbf {\bibinfo {volume} {93}},\ \bibinfo {pages} {024404} (\bibinfo
  {year} {2016})}\BibitemShut {NoStop}%
\bibitem [{\citenamefont {Lau}\ \emph {et~al.}(2017)\citenamefont {Lau},
  \citenamefont {Sheng}, \citenamefont {Mitani}, \citenamefont {Chiba},\ and\
  \citenamefont {Hayashi}}]{lau2017apl}%
  \BibitemOpen
  \bibfield  {author} {\bibinfo {author} {\bibfnamefont {Y.-C.}\ \bibnamefont
  {Lau}}, \bibinfo {author} {\bibfnamefont {P.}~\bibnamefont {Sheng}}, \bibinfo
  {author} {\bibfnamefont {S.}~\bibnamefont {Mitani}}, \bibinfo {author}
  {\bibfnamefont {D.}~\bibnamefont {Chiba}}, \ and\ \bibinfo {author}
  {\bibfnamefont {M.}~\bibnamefont {Hayashi}},\ }\href@noop {} {\bibfield
  {journal} {\bibinfo  {journal} {Applied Physics Letters}\ }\textbf {\bibinfo
  {volume} {110}},\ \bibinfo {pages} {022405} (\bibinfo {year}
  {2017})}\BibitemShut {NoStop}%
\bibitem [{\citenamefont {Pradipto}\ \emph {et~al.}(2019)\citenamefont
  {Pradipto}, \citenamefont {Yakushiji}, \citenamefont {Ham}, \citenamefont
  {Kim}, \citenamefont {Shiota}, \citenamefont {Moriyama}, \citenamefont {Kim},
  \citenamefont {Lee}, \citenamefont {Nakamura}, \citenamefont {Lee},\ and\
  \citenamefont {Ono}}]{pradipto2019prb}%
  \BibitemOpen
  \bibfield  {author} {\bibinfo {author} {\bibfnamefont {A.~M.}\ \bibnamefont
  {Pradipto}}, \bibinfo {author} {\bibfnamefont {K.}~\bibnamefont {Yakushiji}},
  \bibinfo {author} {\bibfnamefont {W.~S.}\ \bibnamefont {Ham}}, \bibinfo
  {author} {\bibfnamefont {S.}~\bibnamefont {Kim}}, \bibinfo {author}
  {\bibfnamefont {Y.}~\bibnamefont {Shiota}}, \bibinfo {author} {\bibfnamefont
  {T.}~\bibnamefont {Moriyama}}, \bibinfo {author} {\bibfnamefont {K.~W.}\
  \bibnamefont {Kim}}, \bibinfo {author} {\bibfnamefont {H.~W.}\ \bibnamefont
  {Lee}}, \bibinfo {author} {\bibfnamefont {K.}~\bibnamefont {Nakamura}},
  \bibinfo {author} {\bibfnamefont {K.~J.}\ \bibnamefont {Lee}}, \ and\
  \bibinfo {author} {\bibfnamefont {T.}~\bibnamefont {Ono}},\ }\href@noop {}
  {\bibfield  {journal} {\bibinfo  {journal} {Phys. Rev. B}\ }\textbf {\bibinfo
  {volume} {99}},\ \bibinfo {pages} {180410} (\bibinfo {year}
  {2019})}\BibitemShut {NoStop}%
\bibitem [{\citenamefont {Chen}\ \emph {et~al.}(1995)\citenamefont {Chen},
  \citenamefont {Idzerda}, \citenamefont {Lin}, \citenamefont {Smith},
  \citenamefont {Meigs}, \citenamefont {Chaban}, \citenamefont {Ho},
  \citenamefont {Pellegrin},\ and\ \citenamefont {Sette}}]{chen1995prl}%
  \BibitemOpen
  \bibfield  {author} {\bibinfo {author} {\bibfnamefont {C.~T.}\ \bibnamefont
  {Chen}}, \bibinfo {author} {\bibfnamefont {Y.~U.}\ \bibnamefont {Idzerda}},
  \bibinfo {author} {\bibfnamefont {H.~J.}\ \bibnamefont {Lin}}, \bibinfo
  {author} {\bibfnamefont {N.~V.}\ \bibnamefont {Smith}}, \bibinfo {author}
  {\bibfnamefont {G.}~\bibnamefont {Meigs}}, \bibinfo {author} {\bibfnamefont
  {E.}~\bibnamefont {Chaban}}, \bibinfo {author} {\bibfnamefont {G.~H.}\
  \bibnamefont {Ho}}, \bibinfo {author} {\bibfnamefont {E.}~\bibnamefont
  {Pellegrin}}, \ and\ \bibinfo {author} {\bibfnamefont {F.}~\bibnamefont
  {Sette}},\ }\href@noop {} {\bibfield  {journal} {\bibinfo  {journal} {Phys.
  Rev. Lett.}\ }\textbf {\bibinfo {volume} {75}},\ \bibinfo {pages} {152}
  (\bibinfo {year} {1995})}\BibitemShut {NoStop}%
\bibitem [{\citenamefont {Kolesnikov}\ \emph {et~al.}(2016)\citenamefont
  {Kolesnikov}, \citenamefont {Stebliy}, \citenamefont {Ognev}, \citenamefont
  {Samardak}, \citenamefont {Fedorets}, \citenamefont {Plotnikov},
  \citenamefont {Han},\ and\ \citenamefont {Chebotkevich}}]{kolesnikov2016jpd}%
  \BibitemOpen
  \bibfield  {author} {\bibinfo {author} {\bibfnamefont {A.~G.}\ \bibnamefont
  {Kolesnikov}}, \bibinfo {author} {\bibfnamefont {M.~E.}\ \bibnamefont
  {Stebliy}}, \bibinfo {author} {\bibfnamefont {A.~V.}\ \bibnamefont {Ognev}},
  \bibinfo {author} {\bibfnamefont {A.~S.}\ \bibnamefont {Samardak}}, \bibinfo
  {author} {\bibfnamefont {A.~N.}\ \bibnamefont {Fedorets}}, \bibinfo {author}
  {\bibfnamefont {V.~S.}\ \bibnamefont {Plotnikov}}, \bibinfo {author}
  {\bibfnamefont {X.~F.}\ \bibnamefont {Han}}, \ and\ \bibinfo {author}
  {\bibfnamefont {L.~A.}\ \bibnamefont {Chebotkevich}},\ }\href@noop {}
  {\bibfield  {journal} {\bibinfo  {journal} {J. Phys. D-Appl. Phys.}\ }\textbf
  {\bibinfo {volume} {49}},\ \bibinfo {pages} {425302} (\bibinfo {year}
  {2016})}\BibitemShut {NoStop}%
\bibitem [{\citenamefont {Grytsyuk}\ \emph {et~al.}(2016)\citenamefont
  {Grytsyuk}, \citenamefont {Belabbes}, \citenamefont {Haney}, \citenamefont
  {Lee}, \citenamefont {Lee}, \citenamefont {Stiles}, \citenamefont
  {Schwingenschlogl},\ and\ \citenamefont {Manchon}}]{grytsyuk2016prb}%
  \BibitemOpen
  \bibfield  {author} {\bibinfo {author} {\bibfnamefont {S.}~\bibnamefont
  {Grytsyuk}}, \bibinfo {author} {\bibfnamefont {A.}~\bibnamefont {Belabbes}},
  \bibinfo {author} {\bibfnamefont {P.~M.}\ \bibnamefont {Haney}}, \bibinfo
  {author} {\bibfnamefont {H.~W.}\ \bibnamefont {Lee}}, \bibinfo {author}
  {\bibfnamefont {K.~J.}\ \bibnamefont {Lee}}, \bibinfo {author} {\bibfnamefont
  {M.~D.}\ \bibnamefont {Stiles}}, \bibinfo {author} {\bibfnamefont
  {U.}~\bibnamefont {Schwingenschlogl}}, \ and\ \bibinfo {author}
  {\bibfnamefont {A.}~\bibnamefont {Manchon}},\ }\href@noop {} {\bibfield
  {journal} {\bibinfo  {journal} {Physical Review B}\ }\textbf {\bibinfo
  {volume} {93}},\ \bibinfo {pages} {174421} (\bibinfo {year}
  {2016})}\BibitemShut {NoStop}%
\bibitem [{\citenamefont {Bruno}(1995)}]{bruno1995prb}%
  \BibitemOpen
  \bibfield  {author} {\bibinfo {author} {\bibfnamefont {P.}~\bibnamefont
  {Bruno}},\ }\href@noop {} {\bibfield  {journal} {\bibinfo  {journal} {Phys.
  Rev. B}\ }\textbf {\bibinfo {volume} {52}},\ \bibinfo {pages} {411} (\bibinfo
  {year} {1995})}\BibitemShut {NoStop}%
\bibitem [{\citenamefont {Stiles}(1999)}]{stiles1999jmmm}%
  \BibitemOpen
  \bibfield  {author} {\bibinfo {author} {\bibfnamefont {M.~D.}\ \bibnamefont
  {Stiles}},\ }\href@noop {} {\bibfield  {journal} {\bibinfo  {journal}
  {Journal of Magnetism and Magnetic Materials}\ }\textbf {\bibinfo {volume}
  {200}},\ \bibinfo {pages} {322} (\bibinfo {year} {1999})}\BibitemShut
  {NoStop}%
\bibitem [{\citenamefont {Hrabec}\ \emph {et~al.}(2014)\citenamefont {Hrabec},
  \citenamefont {Porter}, \citenamefont {Wells}, \citenamefont {Benitez},
  \citenamefont {Burnell}, \citenamefont {McVitie}, \citenamefont {McGrouther},
  \citenamefont {Moore},\ and\ \citenamefont {Marrows}}]{hrabec2014prb}%
  \BibitemOpen
  \bibfield  {author} {\bibinfo {author} {\bibfnamefont {A.}~\bibnamefont
  {Hrabec}}, \bibinfo {author} {\bibfnamefont {N.~A.}\ \bibnamefont {Porter}},
  \bibinfo {author} {\bibfnamefont {A.}~\bibnamefont {Wells}}, \bibinfo
  {author} {\bibfnamefont {M.~J.}\ \bibnamefont {Benitez}}, \bibinfo {author}
  {\bibfnamefont {G.}~\bibnamefont {Burnell}}, \bibinfo {author} {\bibfnamefont
  {S.}~\bibnamefont {McVitie}}, \bibinfo {author} {\bibfnamefont
  {D.}~\bibnamefont {McGrouther}}, \bibinfo {author} {\bibfnamefont {T.~A.}\
  \bibnamefont {Moore}}, \ and\ \bibinfo {author} {\bibfnamefont {C.~H.}\
  \bibnamefont {Marrows}},\ }\href@noop {} {\bibfield  {journal} {\bibinfo
  {journal} {Phys. Rev. B}\ }\textbf {\bibinfo {volume} {90}},\ \bibinfo
  {pages} {020402} (\bibinfo {year} {2014})}\BibitemShut {NoStop}%
\bibitem [{\citenamefont {Je}\ \emph {et~al.}(2013)\citenamefont {Je},
  \citenamefont {Kim}, \citenamefont {Yoo}, \citenamefont {Min}, \citenamefont
  {Lee},\ and\ \citenamefont {Choe}}]{je2013prb}%
  \BibitemOpen
  \bibfield  {author} {\bibinfo {author} {\bibfnamefont {S.-G.}\ \bibnamefont
  {Je}}, \bibinfo {author} {\bibfnamefont {D.-H.}\ \bibnamefont {Kim}},
  \bibinfo {author} {\bibfnamefont {S.-C.}\ \bibnamefont {Yoo}}, \bibinfo
  {author} {\bibfnamefont {B.-C.}\ \bibnamefont {Min}}, \bibinfo {author}
  {\bibfnamefont {K.-J.}\ \bibnamefont {Lee}}, \ and\ \bibinfo {author}
  {\bibfnamefont {S.-B.}\ \bibnamefont {Choe}},\ }\href@noop {} {\bibfield
  {journal} {\bibinfo  {journal} {Phys. Rev. B}\ }\textbf {\bibinfo {volume}
  {88}},\ \bibinfo {pages} {214401} (\bibinfo {year} {2013})}\BibitemShut
  {NoStop}%
\bibitem [{\citenamefont {Wells}\ \emph {et~al.}(2017)\citenamefont {Wells},
  \citenamefont {Shepley}, \citenamefont {Marrows},\ and\ \citenamefont
  {Moore}}]{wells2017prb}%
  \BibitemOpen
  \bibfield  {author} {\bibinfo {author} {\bibfnamefont {A.~W.~J.}\
  \bibnamefont {Wells}}, \bibinfo {author} {\bibfnamefont {P.~M.}\ \bibnamefont
  {Shepley}}, \bibinfo {author} {\bibfnamefont {C.~H.}\ \bibnamefont
  {Marrows}}, \ and\ \bibinfo {author} {\bibfnamefont {T.~A.}\ \bibnamefont
  {Moore}},\ }\href@noop {} {\bibfield  {journal} {\bibinfo  {journal} {Phys.
  Rev. B}\ }\textbf {\bibinfo {volume} {95}},\ \bibinfo {pages} {054428}
  (\bibinfo {year} {2017})}\BibitemShut {NoStop}%
\bibitem [{\citenamefont {Kim}\ \emph {et~al.}(2016{\natexlab{c}})\citenamefont
  {Kim}, \citenamefont {Song}, \citenamefont {Choi}, \citenamefont {Min},
  \citenamefont {Kim}, \citenamefont {Choi},\ and\ \citenamefont
  {Lee}}]{kim2016scirep}%
  \BibitemOpen
  \bibfield  {author} {\bibinfo {author} {\bibfnamefont {D.~O.}\ \bibnamefont
  {Kim}}, \bibinfo {author} {\bibfnamefont {K.~M.}\ \bibnamefont {Song}},
  \bibinfo {author} {\bibfnamefont {Y.}~\bibnamefont {Choi}}, \bibinfo {author}
  {\bibfnamefont {B.~C.}\ \bibnamefont {Min}}, \bibinfo {author} {\bibfnamefont
  {J.~S.}\ \bibnamefont {Kim}}, \bibinfo {author} {\bibfnamefont {J.~W.}\
  \bibnamefont {Choi}}, \ and\ \bibinfo {author} {\bibfnamefont {D.~R.}\
  \bibnamefont {Lee}},\ }\href@noop {} {\bibfield  {journal} {\bibinfo
  {journal} {Scientific Reports}\ }\textbf {\bibinfo {volume} {6}},\ \bibinfo
  {pages} {25391} (\bibinfo {year} {2016}{\natexlab{c}})}\BibitemShut {NoStop}%
\end{thebibliography}%

\end{document}